\documentclass[english,final]{IEEEtran}
\usepackage[T1]{fontenc}
\usepackage[latin9]{inputenc}
\usepackage{amsthm}
\usepackage{amsmath}
\usepackage{amssymb}
\usepackage{graphicx}
\usepackage{esint}
\usepackage{epstopdf}

\makeatletter
\theoremstyle{plain}
\newtheorem{thm}{\protect\theoremname}
\newtheorem{lem}{\protect\lemmaname}

\usepackage{cite}
\usepackage{enumerate}
\usepackage{amsmath}
\usepackage{flushend}
\usepackage{dsfont}

\newcommand{\openone}{\mathds{1}}

\newcommand{\bPhi}{\boldsymbol{\Phi}}
\newcommand{\fvrem}{\boldsymbol{f}_{\mathrm{rem}}}
\newcommand{\Vvf}{\boldsymbol{V}_{\hspace*{-0.3ex}\boldsymbol{f}}}

\newcommand{\SetScc}{\mathcal{S}}

\newcommand{\Ptilde}{\widetilde{P}}
\newcommand{\Qtilde}{\widetilde{Q}}

\newcommand{\uv}{\boldsymbol{u}}

\newcommand{\Uv}{\boldsymbol{U}}

\newcommand{\Vv}{\boldsymbol{V}}

\newcommand{\xv}{\boldsymbol{x}}

\newcommand{\Ytilde}{\widetilde{Y}}

\newcommand{\Xv}{\boldsymbol{X}}
\newcommand{\yv}{\boldsymbol{y}}
\newcommand{\Yv}{\boldsymbol{Y}}
\newcommand{\zv}{\boldsymbol{z}}
\newcommand{\Zv}{\boldsymbol{Z}}

\newcommand{\Vvi}{\boldsymbol{V}_{\hspace*{-0.3ex}1}} 
\newcommand{\Vvii}{\boldsymbol{V}_{\hspace*{-0.3ex}2}}
\newcommand{\Vvm}{\boldsymbol{V}_{\hspace*{-0.3ex}m}}

\newcommand{\Ac}{\mathcal{A}}

\newcommand{\Cc}{\mathcal{C}}

\newcommand{\Kc}{\mathcal{K}}
\newcommand{\Pc}{\mathcal{P}}

\newcommand{\Uc}{\mathcal{U}}

\newcommand{\Xc}{\mathcal{X}}
\newcommand{\Yc}{\mathcal{Y}}

\newcommand{\EE}{\mathbb{E}}
\newcommand{\PP}{\mathbb{P}}
\newcommand{\RR}{\mathbb{R}}

\newcommand{\Qsf}{\mathsf{Q}}

\newcommand{\defeq}{\triangleq}
\newcommand{\deq}{\stackrel{d}{=}}
\newcommand{\var}{\mathrm{Var}}
\newcommand{\cov}{\mathrm{Cov}}

\newcommand{\Av}{\boldsymbol{A}}
\newcommand{\Bv}{\boldsymbol{B}}
\newcommand{\bv}{\boldsymbol{b}}

\newcommand{\csf}{\mathsf{c}}

\newcommand{\Rc}{\mathcal{R}}

\newcommand{\Lc}{\mathcal{L}}

\newcommand{\Iv}{\boldsymbol{I}}
\newcommand{\iv}{\boldsymbol{i}}
\newcommand{\Fv}{\boldsymbol{F}}
\newcommand{\fv}{\boldsymbol{f}}

\newcommand{\Lv}{\boldsymbol{L}}
\newcommand{\Rv}{\boldsymbol{R}}

\newcommand{\Sv}{\boldsymbol{S}}
\newcommand{\Tv}{\boldsymbol{T}}
\newcommand{\bzero}{\boldsymbol{0}}
\newcommand{\bone}{\boldsymbol{1}}
\newcommand{\bmu}{\boldsymbol{\mu}}

\newcommand{\bgamma}{\boldsymbol{\gamma}}

\newcommand{\bOmega}{\boldsymbol{\Omega}}
\newcommand{\bSigma}{\boldsymbol{\Sigma}}
\newcommand{\bLambda}{\boldsymbol{\Lambda}}

\newcommand{\bDelta}{\boldsymbol{\Delta}}

\newcommand{\Qinv}{\mathsf{Q}_{\mathrm{inv}}}

\@ifundefined{showcaptionsetup}{}{%
 \PassOptionsToPackage{caption=false}{subfig}}
\usepackage{subfig}
\makeatother

\usepackage{babel}
\providecommand{\theoremname}{Theorem}
\providecommand{\lemmaname}{Lemma}

\begin{document}

\title{Second-Order Rate Region of Constant-Composition Codes for the Multiple-Access
Channel}

\author{Jonathan Scarlett, Alfonso Martinez and Albert Guill\'en i F\`abregas}

\maketitle
\long\def\symbolfootnote[#1]#2{\begingroup\def\thefootnote{\fnsymbol{footnote}}\footnote[#1]{#2}\endgroup}
\begin{abstract}
    This paper studies the second-order asymptotics of coding rates for
    the discrete memoryless multiple-access channel with a fixed target
    error probability. Using constant-composition random coding, coded
    time-sharing, and a variant of Hoeffding's combinatorial central limit
    theorem, an inner bound on the set of locally achievable second-order
    coding rates is given for each point on the boundary of the capacity
    region. It is shown that the inner bound for constant-composition
    random coding includes that recovered by i.i.d.~random coding, and
    that the inclusion may be strict. The inner bound is extended to the
    Gaussian multiple-access channel via an increasingly fine quantization
    of the inputs.
\end{abstract}

\symbolfootnote[0]{J. Scarlett was with the Department of Engineering, University 
of Cambridge, Cambridge, CB2 1PZ, U.K.  He is now with the Laboratory
for Information and Inference Systems, \'Ecole Polytechnique F\'ed\'erale
de Lausanne, CH-1015, Switzerland (e-mail: jmscarlett@gmail.com).  

A. Martinez is with the Department of Information and Communication Technologies, 
Universitat Pompeu Fabra, 08018 Barcelona, Spain (e-mail: alfonso.martinez@ieee.org).  

A. Guill\'en i F\`abregas is with the Instituci\'o Catalana de Recerca i Estudis 
Avan\c{c}ats (ICREA), the Department of Information and Communication Technologies, 
Universitat Pompeu Fabra, 08018 Barcelona, Spain, and also with the Department of 
Engineering, University of Cambridge, Cambridge, CB2 1PZ, U.K. (e-mail: 
guillen@ieee.org).

This work has been funded in part by the European Research Council under ERC 
grant agreement 259663, by the European Union's 7th Framework Programme under 
grant agreement 303633 and by the Spanish Ministry of Economy and Competitiveness 
under grants RYC-2011-08150 and TEC2012-38800-C03-03. This work was presented in 
part at the 51st Allerton Conference on Communication, Computing and Control (2013).}

\section{Introduction \label{sec:INTRO}}

The channel capacity describes the highest rate of transmission
with vanishing error probability in coded communication systems. Further
characterizations of the system performance are given by error exponents
\cite[Ch. 9]{FanoBook}, moderate deviations results \cite{ModerateDev},
and second-order coding rates \cite{Strassen}. The latter has regained
significant attention in recent years \cite{Finite,Hayashi}, and
is well-understood for a variety of settings. For discrete
memoryless channels, the maximum number of codewords of length $n$ 
yielding an error probability not exceeding $\epsilon\in(0,1)$,
denoted by $M^{*}(n,\epsilon)$, satisfies \cite{Strassen}
\begin{equation}
    \log M^{*}(n,\epsilon)=nC-\sqrt{nV}\Qsf^{-1}(\epsilon)+o(\sqrt{n}),\label{eq:MD_SingleUser0}
\end{equation}
where $C$ is the channel capacity, $\Qsf^{-1}(\cdot)$ is the functional
inverse of the standard Gaussian tail probability $\Qsf(z)\defeq\int_{z}^{\infty}\frac{1}{\sqrt{2\pi}}e^{-\frac{z^{2}}{2}}dz$,
and $V$ is known as the channel dispersion. Expansions of the form
\eqref{eq:MD_SingleUser0} provide additional insight into the system
performance beyond the capacity alone by quantifying the rate of convergence.

In this paper, we study the second-order asymptotics of the multiple-access
channel (MAC). Achievability results for this problem have previously
been obtained using i.i.d.~random coding with a random time-sharing
sequence \cite{MACFinite1,MACFinite2} and a deterministic time-sharing
sequence \cite{MACFinite3}, whereas we demonstrate improved
asymptotic bounds via the use of constant-composition random coding
\cite[Ch. 9]{FanoBook}. A key tool in our analysis is a Berry-Esseen
theorem associated with a variant of Hoeffding's combinatorial central
limit theorem (CLT) \cite{CombHoeffding}.  We consider a 
local notion of second-order achievability proposed by
Nomura and Han \cite{NomuraHan}, in which the second-order coding rates
(e.g.~$-\sqrt{V}\Qsf^{-1}(\epsilon)$ in \eqref{eq:MD_SingleUser0})
of the users are sought for a fixed point on the boundary of the capacity
region.

\subsection{\label{sub:MD_NOTATION}Notation}

The set of all probability distributions on an alphabet $\Xc$
is denoted by $\Pc(\Xc)$, and the set of conditional
distributions on $\Yc$ given $\Xc$ is denoted by
$\Pc(\Yc|\Xc)$. Given a distribution $Q(x)$
and a conditional distribution $W(y|x)$, the joint distribution $Q(x)W(y|x)$
is denoted by $Q\times W$. The set of all empirical distributions
(i.e.~types \cite[Ch. 2]{CsiszarBook}) for sequences in $\Xc^{n}$
is denoted by $\Pc_{n}(\Xc)$. The set of all sequences
of length $n$ with a given type $P_{X}$ is denoted by $T^{n}(P_{X})$,
and similarly for joint types. Given a sequence $\xv\in T^{n}(P_{X})$
and a conditional distribution $P_{Y|X}$, we define $T_{\xv}^{n}(P_{Y|X})$
to be the set of sequences $\yv$ such that $(\xv,\yv)\in T^{n}(P_{X}\times P_{Y|X})$.

Bold symbols are used for vectors and matrices (e.g.~$\xv$),
and the corresponding $i$-th entry of a vector is written using a
subscript (e.g.~$x_{i}$). The vectors (or matrices) of all zeros
and all ones are denoted by $\bzero$ and $\bone$
respectively, and the identity matrix is denoted by $\mathbb{I}$;
the sizes will be clear from the context. The symbols $\prec$, $\preceq$,
etc.~denote element-wise inequalities for vectors, and inequalities
on the positive semidefinite cone for matrices (e.g.~$\Vv\succ\bzero$
means $\Vv$ is positive definite). We denote the $\ell_{2}$-norm
of a vector by $\|\cdot\|$, and the maximum absolute value of the
entries of a vector or matrix by $\|\cdot\|_{\infty}$. We denote
the transpose of a vector or matrix by $(\cdot)^{T}$, the inverse
of a matrix by $(\cdot)^{-1}$, the positive definite matrix square
root by $(\cdot)^{\frac{1}{2}}$, and its inverse by $(\cdot)^{-\frac{1}{2}}$.
The multivariate Gaussian distribution with mean $\bmu$
and covariance matrix $\bSigma$ is denoted by $N(\bmu,\bSigma)$.

We denote the cross-covariance matrix of two random vectors by $\cov[\Zv_{1},\Zv_{2}]=\EE\big[(\Zv_{1}-\EE[\Zv_{1}])(\Zv_{2}-\EE[\Zv_{2}])^{T}\big]$,
and we write $\cov[\Zv]$ in place of $\cov[\Zv,\Zv]$.
The variance of a scalar random variable is denoted by $\var[\cdot]$.
Logarithms have base $e$, and all rates are in nats except in the
examples, where bits are used. We denote the indicator function by
$\openone\{\cdot\}$. For a set $\SetScc$ of real numbers (or
vectors) and a constant (or vector) $c$, we write $\SetScc+c$
(or $c+\SetScc$) to denote the set $\{s+c\,:\, s\in\SetScc\}$.
We similarly write $c\SetScc=\{cs\,:\, s\in\SetScc\}$ for
a given constant $c$.

For two sequences $f_{n}$ and $g_{n}$, we write $f_{n}=O(g_{n})$
if $|f_{n}|\le c|g_{n}|$ for some $c$ and sufficiently large $n$,
and $f_{n}=o(g_{n})$ if $\lim_{n\to\infty}\frac{f_{n}}{g_{n}}=0$.
We write $f_{n}=\Theta(g_{n})$ if $f_{n}=O(g_{n})$ and $g_{n}=O(f_{n})$.

\subsection{\label{sub:MD_SYSTEM_MODEL}System Setup and Definitions}

We consider a two-user discrete memoryless MAC (DM-MAC) $W(y|x_{1},x_{2})$
with input alphabets $\Xc_{1}$ and $\Xc_{2}$ and
output alphabet $\Yc$, yielding an $n$-letter transition
law given by $W^{n}(\yv|\xv_{1},\xv_{2})\defeq\prod_{i=1}^{n}W(y_{i}|x_{1,i},x_{2,i})$.
The encoders and decoder operate as follows. Encoder $\nu=1,2$ takes
as input a message $m_{\nu}$ equiprobable on the set $\{1,\dotsc,M_{\nu}\}$,
and transmits the corresponding codeword $\xv_{\nu}^{(m_{\nu})}$
from the codebook $\Cc_{\nu}=\{\xv_{\nu}^{(1)},\dotsc,\xv_{\nu}^{(M_{\nu})}\}$.
The decoder forms an estimate $(\hat{m}_{1},\hat{m}_{2})$ of the
message pair using the output sequence $\yv$ and the two
codebooks. An error is said to have occurred if 
$(\hat{m}_{1},\hat{m}_{2}) \ne (m_{1},m_{2})$. A rate pair $(R_{1},R_{2})$ is said
to be $(n,\epsilon)$-achievable if there exist codebooks with $M_{1}\ge e^{nR_{1}}$
and $M_{2}\ge e^{nR_{2}}$ codewords of length $n$ for users 1 and
2 respectively, such that the average error probability does not exceed
$\epsilon$. The capacity region $\Rc^{*}$ is defined to
be the closure of the set of rate pairs $(R_{1},R_{2})$ that are
$(n,\epsilon)$-achievable for any $\epsilon\in(0,1)$ and sufficiently
large $n$.

Our results are proved using constant-composition random coding with
coded time-sharing \cite{MACExponent4}. The precise description of
the ensemble is postponed until Section \ref{sec:MD_PROOF}; here
we simply provide the definitions required to state the results. We
fix a finite time-sharing alphabet $\Uc$, as well as the
input distributions $Q_{U}(u)$, $Q_{1}(x_{1}|u)$ and $Q_{2}(x_{2}|u)$.
We define the joint distribution
\begin{multline}
    P_{UX_{1}X_{2}Y}(u,x_{1},x_{2},y) \\ \defeq Q_{U}(u)Q_{1}(x_{1}|u)Q_{2}(x_{2}|u)W(y|x_{1},x_{2}),\label{eq:MD_Distr}
\end{multline}
and denote the induced marginal distributions by $P_{Y|X_{1}U}$,
$P_{Y|U}$, etc. Defining the rate vector
\begin{equation}
    \Rv\defeq\left[\begin{array}{c}
        R_{1}\\
        R_{2}\\
        R_{1}+R_{2}
    \end{array}\right]\label{eq:MD_VecR}
\end{equation}
and the mutual information vector (implicitly dependent 
on $Q_U$, $Q_1$, $Q_2$ and $W$)
\begin{equation}
    \Iv\defeq\left[\begin{array}{c}
        I(X_{1};Y|X_{2},U)\\
        I(X_{2};Y|X_{1},U)\\
        I(X_{1},X_{2};Y|U)
    \end{array}\right],\label{eq:MD_VecI}
\end{equation}
we have \cite{MACCapacity1,MACCapacity2,NetworkBook}
\begin{equation}
    \Rc^{*}=\bigcup_{\Uc}\bigcup_{Q_{U},Q_{1},Q_{2}}\Big\{(R_{1},R_{2})\,:\,\Rv\preceq\Iv\Big\}.\label{eq:MD_CapRegion}
\end{equation}
Moreover, the union over $\Uc$ may be restricted to satisfy
$|\,\Uc|\le2$. The three conditions in the element-wise inequality
$\Rv\preceq\Iv$ correspond to a treatment of
the error event as a union of three error types:

\begin{tabbing}     
    ~~~{\emph{(Type 1)}}~~~ \= $\hat{m}_1 = m_1$ and $\hat{m}_2 \ne m_2$, \\
    ~~~{\emph{(Type 2)}}~~~ \> $\hat{m}_1 \ne m_1$ and $\hat{m}_2 = m_2$, \\   
    ~~~{\emph{(Type 12)}}~~~ \> $\hat{m}_1 \ne m_1$ and $\hat{m}_2 \ne m_2$.
\end{tabbing}

A key quantity in our analysis is the information density vector \cite{MACFinite1,MACFinite3}
\begin{equation}
    \iv(u,x_{1},x_{2},y)\defeq\left[\begin{array}{c}
        i_{1}(u,x_{1},x_{2},y)\\
        i_{2}(u,x_{1},x_{2},y)\\
        i_{12}(u,x_{1},x_{2},y)
    \end{array}\right],\label{eq:MD_bold_i}
\end{equation}
where
\begin{align}
    i_{1}(u,x_{1},x_{2},y)  &\defeq\log\frac{W(y|x_{1},x_{2})}{P_{Y|X_{2}U}(y|x_{2},u)}\label{eq:MD_i1}\\
    i_{2}(u,x_{1},x_{2},y)  &\defeq\log\frac{W(y|x_{1},x_{2})}{P_{Y|X_{1}U}(y|x_{1},u)}\\
    i_{12}(u,x_{1},x_{2},y) &\defeq\log\frac{W(y|x_{1},x_{2})}{P_{Y|U}(y|u)}.\label{eq:MD_i12}
\end{align}
Averaging $\iv$ with respect to the distribution in \eqref{eq:MD_Distr}
yields the mutual information vector in \eqref{eq:MD_VecI}.

We consider the local notion of second-order asymptotics introduced
by Nomura and Han for the Slepian-Wolf problem \cite{NomuraHan}; see
also Hayashi \cite{Hayashi} for the analogous definitions in the
single-user setting. We proceed by presenting similar definitions
for the present setting, albeit in a slightly different form. A pair
$(L_{1},L_{2})$ is said to be $(n,\epsilon,R_{1}^{*},R_{2}^{*})$-achievable
if there exist codebooks with $M_{\nu}\geq e^{nR_{\nu}^{*}+\sqrt{n}L_{\nu}}$
codewords of length $n$ for $\nu=1,2$ such that the average error
probability $p_{e}$ does not exceed $\epsilon$. The \emph{second-order
    rate region} \emph{$\Lc(\epsilon,R_{1}^{*},R_{2}^{*})$} is
defined as the closure of the set of pairs $(L_{1},L_{2})$ that are
$(n,\epsilon,R_{1}^{*},R_{2}^{*})$-achievable for sufficiently large
$n$. In other words, $\Lc(\epsilon,R_{1}^{*},R_{2}^{*})$
is the set of all $(L_{1},L_{2})$ pairs for which there exists an
$\epsilon$-reliable code with $\log M_{\nu}=nR_{\nu}^{*}+\sqrt{n}L_{\nu}+o(\sqrt{n})$
for $\nu=1,2$. While this definition is valid for any pair $(R_{1}^{*},R_{2}^{*})$,
our focus will be on pairs on the boundary of the capacity region
$\Rc^{*}$; in all other cases we trivially have either $\Lc=\emptyset$
or $\Lc=\RR^{2}$.  We will see that both negative and positive values
of $L_{\nu}$ arise; the former can be thought of as a backoff
from the first-order term, and the latter an addition to the 
first-order term.

Finally, we define the set
\begin{equation}
    \Qsf_{\mathrm{inv}}(\Vv,\epsilon)\defeq\Big\{\zv\in\RR^{d}\,:\,\PP\big[\Zv\preceq\zv]\ge1-\epsilon\Big\},
\end{equation}
where $\Zv\sim N(\bzero,\Vv)$, and
$\Vv$ is a $d\times d$ positive semi-definite matrix.
This definition applies for an arbitrary dimension $d$, which is
dictated by the first argument.

\subsection{\label{sub:MD_EXISTING_RESULTS} Previous Work}

The second-order rate region $\Lc$ has been characterized
for very few multi-user problems \cite{NomuraHan,PaperCMAC,DispIC}.
The one most relevant to this paper is the Gaussian MAC with degraded
message sets \cite{PaperCMAC}, which has the notable feature of having
a curved capacity region, giving rise to a non-standard derivative
term in the expression for $\Lc$. Our analysis will yield
similar terms using different techniques.

Tan and Kosut \cite{MACFinite1} and Haim \emph{et al.} \cite{MACFinite5}
performed second-order asymptotic studies for various multi-user problems using
different  notions of achievability to those above.  In particular, both
of these works considered the problem of finding the backoff from
the rates when a point $(R_{1}^{*},R_{2}^{*})$ is approached from
a given angle. As demonstrated in \cite{PaperCMAC}, this problem
can be solved numerically in a straightforward fashion once 
$\Lc(\epsilon,R_{1}^{*},R_{2}^{*})$ is characterized. 

Other previous works on the DM-MAC have taken an 
alternative approach to characterizing the second-order asymptotics, namely
seeking \emph{global} asymptotic expansions of the following form:
For any triplet $(Q_{U},Q_{1},Q_{2})$, rate vectors $\Rv$
satisfying 
\begin{equation}
    n\Rv\in n\Iv-\sqrt{n}\,\Qsf_{\mathrm{inv}}(\Vv,\epsilon)+g(n)\bone,\label{eq:MD_Achievable}
\end{equation}
are $(n,\epsilon)$-achievable for some \emph{dispersion matrix} $\Vv$
and function $g(n)=o(\sqrt{n})$, where $\Iv$ is given
in \eqref{eq:MD_VecI}.

The first global result for the DM-MAC was given in \cite{MACFinite1},
where i.i.d.~random coding was used to obtain \eqref{eq:MD_Achievable}
with $\Iv=\EE[\iv(U,X_{1},X_{2},Y)]$ and $\Vv=\cov[\iv(U,X_{1},X_{2},Y)]$
(see also \cite{MACFinite2}).  Expansions of a similar form were given
by MolavianJazi and Laneman \cite{MACFinite2}.  By using a 
constant-composition time-sharing sequence, Huang and Moulin \cite{MACFinite3}
showed that the dispersion matrix can be improved to
\begin{equation}
    \Vv^{\mathrm{iid}}\defeq\EE\Big[\cov\big[\iv(U,X_{1},X_{2},Y)\,\big|\, U\big]\Big].\label{eq:MD_V_iid}
\end{equation}

As discussed by Haim \emph{et al. }\cite{MACFinite5}, expansions
of the form \eqref{eq:MD_Achievable} are more difficult to interpret
than the scalar counterpart in \eqref{eq:MD_SingleUser0}, as the 
notion of the convergence of a region is inherently less concrete
than that of the convergence of a scalar. While the
scalar dispersion $V$ in \eqref{eq:MD_SingleUser0} corresponds to
a concrete operational definition \cite{Finite}, it appears difficult
to directly give any such meaning to the matrix $\Vv$ based on global results.
In fact, non-global asymptotic studies in \cite{MACFinite1} indicate
that, in most cases of interest, entries $(1,2)$ and $(2,1)$ of
the matrix do not play a fundamental role in characterizing the performance.
Furthermore, it may be difficult to compare two dispersion matrices,
since the partial positive definite ordering does not guarantee that
at least one of $\Vvi\preceq\Vvii$ or
$\Vvii\preceq\Vvi$ hold. These issues
are even more troublesome when one considers the union over
all input distributions; for example, standard proofs often
yield a non-uniform remainder term $g(n)$ in \eqref{eq:MD_Achievable}.
These limitations motivate the study of local asymptotics, such as
\emph{$\Lc(\epsilon,R_{1}^{*},R_{2}^{*})$} defined above.  However,
global results often prove useful as an intermediate
step towards the local results.

\subsection{Contributions}

The main result of this paper is an inner bound on $\Lc(\epsilon,R_{1}^{*},R_{2}^{*})$
for the discrete memoryless MAC. The result is proved using constant-composition
random coding and a variant of Hoeffding's combinatorial CLT \cite{CombHoeffding,CombLatin}.
Since coding with fixed input distributions (not varying with $n$) is not sufficient to 
achieve all $(L_1,L_2)$ pairs in network information theory problems \cite{PaperCMAC}, 
we apply coded time-sharing \cite[Sec.~4.5.3]{NetworkBook} between input distributions
corresponding to two points on the boundary of the capacity region,
with one of the points only corresponding to a fraction $O\big(\frac{1}{\sqrt{n}}\big)$
of the block length. Several examples are provided, including (i)
a case where constant-composition random coding yields a strictly
larger inner bound than that of i.i.d.~random coding, and (ii) an application to the Gaussian
MAC via a quantization argument.

\section{Main Result \label{sec:MD_MAIN_RESULT}}

\subsection{Further Definitions \label{sub:MD_PRELIM}}

Our main result is written in terms of a dispersion matrix of the
form 
\begin{align}
    \Vv & \defeq\EE\Big[\cov\big[\iv(U,X_{1},X_{2},Y)\,\big|\, U\big] \nonumber \\
    & \quad -\cov\big[\iv^{(1)}(U,X_{1})\,\big|\, U\big]-\cov\big[\iv^{(2)}(U,X_{2})\,\big|\, U\big]\Big],\label{eq:MD_CovMatrix} 
\end{align}
where
\begin{align}
    \iv^{(1)}(u,x_{1}) & \defeq\EE\big[\iv(U,X_{1},X_{2},Y)\,\big|\,(U,X_{1})=(u,x_{1})\big]\label{eq:MD_ivec_1}  \\
    \iv^{(2)}(u,x_{2}) & \defeq\EE\big[\iv(U,X_{1},X_{2},Y)\,\big|\,(U,X_{2})=(u,x_{2})\big].\label{eq:MD_ivec_2} 
\end{align}
We can interpret \eqref{eq:MD_CovMatrix} as follows: The term $\cov[\iv]$
represents the variations in $(X_{1},X_{2},Y)$ in the i.i.d.~case
(cf. \eqref{eq:MD_V_iid}), and the terms $\cov[\iv^{(1)}]$
and $\cov[\iv^{(2)}]$ represent the reduced variations
in $X_{1}$ and $X_{2}$ respectively, resulting from the codewords
having a fixed composition. Since all covariance matrices are positive
semidefinite, we clearly have $\Vv\preceq\Vv^{\mathrm{iid}}$.
We henceforth write the entries of $\Iv$ (see \eqref{eq:MD_VecI})
and $\Vv$ using subscripts: 
\begin{equation}
    \Iv=\left[\begin{array}{c}
        I_{1}\\
        I_{2}\\
        I_{12}
        
        \end{array}\right],\quad\Vv=\left[\begin{array}{ccc}
        V_{1} & V_{1,2} & V_{1,12}\\
        V_{1,2} & V_{2} & V_{2,12}\\
        V_{1,12} & V_{2,12} & V_{12}
    \end{array}\right].
\end{equation}

\begin{figure}
    \begin{centering}
        \subfloat[Pentagonal region]{\begin{centering}
            \includegraphics[width=0.95\columnwidth]{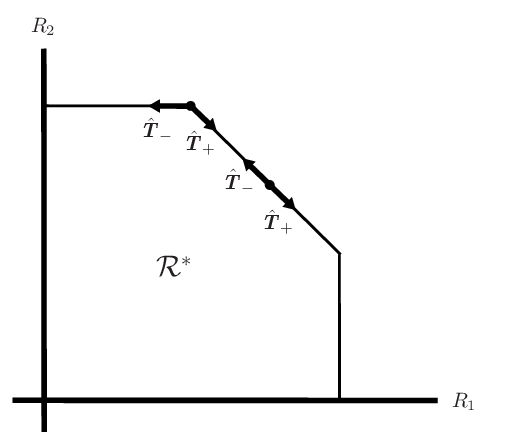}
            \par
        \end{centering}}
        
        \subfloat[Curved region]{\begin{centering}
        \includegraphics[width=0.95\columnwidth]{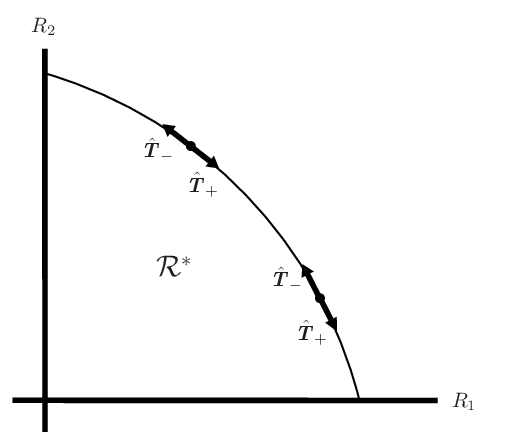}
        \par
        \end{centering}} 
    \par
    \end{centering}
    
    \begin{centering}
        \protect\caption{Illustration of the vectors $\hat{\Tv}_{-}$ and $\hat{\Tv}_{+}$
            for various boundary points of two hypothetical capacity regions.
            \label{fig:MD_ExamplesT}}
        
        \par
    \end{centering}
    
\end{figure}

For a fixed point $(R_{1}^{*},R_{2}^{*})$ on the boundary of $\Rc^{*}$,
we let $\hat{\Tv}_{-}\defeq\hat{\Tv}_{-}(R_{1}^{*},R_{2}^{*})$
and $\hat{\Tv}_{+}\defeq\hat{\Tv}_{+}(R_{1}^{*},R_{2}^{*})$
respectively denote the left and right unit tangent vectors along
the boundary of $\Rc^{*}$ in $(R_{1},R_{2})$-space; see
Figure \ref{fig:MD_ExamplesT}.  We let $\hat{\Tv}_{-}$
(respectively, $\hat{\Tv}_{+}$) be undefined when $R_{1}^{*}=0$
(respectively, $R_{2}^{*}=0$); in all other cases, the vectors are
well-defined due to the convexity of the capacity region. The
case $\hat{\Tv}_{-}=-\hat{\Tv}_{+}$ corresponds
to a curved or straight-line part of the boundary, whereas $\hat{\Tv}_{-}\ne-\hat{\Tv}_{+}$
corresponds to a sudden change in slope (e.g.~at a corner point).

We construct the following vectors in the same way as \eqref{eq:MD_VecR}:
\begin{align}
    \Tv_{-} & \defeq\left[\begin{array}{c} 
    \hat{T}_{-,1}\\
    \hat{T}_{-,2}\\
    \hat{T}_{-,1}+\hat{T}_{-,2}
\end{array}\right],\qquad\Tv_{+}\defeq\left[\begin{array}{c}
    \hat{T}_{+,1}\\
    \hat{T}_{+,2}\\
    \hat{T}_{+,1}+\hat{T}_{+,2}
\end{array}\right],\label{eq:MD_VecsT}
\end{align}
where $\hat{T}_{(\cdot),i}$ denotes the $i$-th entry of the corresponding
unit tangent vector. To ease some of the subsequent discussions, we
define the following scalars that correspond to $\hat{\Tv}_{-}$
and $\hat{\Tv}_{+}$ in a one-to-one fashion:
\begin{equation}
    D_{-}\defeq\frac{\hat{T}_{-,2}}{\hat{T}_{-,1}},\qquad D_{+}\defeq\frac{\hat{T}_{+,2}}{\hat{T}_{+,1}}.\label{eq:MD_Ds}
\end{equation}
These are the left and right derivatives of $R_{2}^{*}$ as a function
of $R_{1}^{*}$. They are non-positive, and are understood to equal
$-\infty$ when $\hat{T}_{(\cdot),1}=0$, corresponding to a vertical
part of the capacity region. Observe that $\hat{\Tv}_{-}$
is obtained by normalizing $[-1\,\,\,{-D_{-}}]$ and $\hat{\Tv}_{+}$
is obtained by normalizing $[1\,\,\, D_{+}]$, and hence $\hat{\Tv}_{-}=-\hat{\Tv}_{+}$
if and only if $D_{-}=D_{+}$.

Given the pairs $(R_{1}^{*},R_{2}^{*})$ and $(L_{1},L_{2})$, we
define
\begin{equation}
    \Rv^{*}\defeq\left[\begin{array}{c}
        R_{1}^{*}\\
        R_{2}^{*}\\
        R_{1}^{*}+R_{2}^{*}
        
        \end{array}\right],\qquad\Lv\defeq\left[\begin{array}{c}
        L_{1}\\
        L_{2}\\
        L_{1}+L_{2}
    \end{array}\right].
\end{equation}
For a non-empty index set $\Kc\subseteq\{1,2,12\}$, we let
$\Lv^{(\Kc)}$ denote the subvector of $\Lv$
where only the indices corresponding to $\Kc$ are kept, and
similarly for $\Rv^{(\Kc)}$, $\Rv^{*(\Kc)}$,
$\Tv_{-}^{(\Kc)}$, $\Tv_{+}^{(\Kc)}$
and $\Iv^{(\Kc)}$. Similarly, $\Vv^{(\Kc)}$
denotes the $|\Kc|\times|\Kc|$ submatrix of $\Vv$
where only the rows and columns indexed by $\Kc$ are kept.

\subsection{Statement of Main Result} \label{sec:MD_STATEMENT}

We say that the triplet $(Q_{U},Q_{1},Q_{2})$
achieves the rate pair $(R_{1},R_{2})$ if $\Rv\preceq\Iv$;
from \eqref{eq:MD_CapRegion}, every point in
$\Rc^*$ (including those on the boundary) is achieved by at 
least one such triplet. In the following theorem, $\Kc$ can be thought of as the
set of error types that are active for a given input distribution
and boundary point (e.g.~if the boundary point is achieved by the
corner point of the pentagonal region corresponding to the type-2
and type-12 conditions, then $\Kc=\{2,12\}$). 
\begin{thm}
    \textup{\emph{\label{thm:MD_LocalResult}Fix $\epsilon\in(0,1)$,
        let $(R_{1}^{*},R_{2}^{*})$ be a point on the boundary of the capacity
        region $\Rc^{*}$ in \eqref{eq:MD_CapRegion}, let $(Q_{U},Q_{1},Q_{2})$
        be an arbitrary triplet achieving that point, and consider $\Iv$
        and $\Vv$ in }}\eqref{eq:MD_VecI} and \eqref{eq:MD_CovMatrix}\textup{\emph{
        respectively. }}Letting $\Kc\subseteq\{1,2,12\}$ be the set
of indices of the largest cardinality such that $\Rv^{*(\Kc)}=\Iv^{(\Kc)}$,
we have 
\begin{align}
    &\Lc(\epsilon,R_{1}^{*},R_{2}^{*}) \nonumber \\
        & \supseteq\bigg\{(L_{1},L_{2})\,:\,\Lv^{(\Kc)}\in\bigcup_{\beta\ge0}\Big\{\beta\Tv_{-}^{(\Kc)}-\Qsf_{\mathrm{inv}}(\Vv^{(\Kc)},\epsilon)\Big\}\bigg\}\nonumber                \\
        & \quad\cup\bigg\{(L_{1},L_{2})\,:\,\Lv^{(\Kc)}\in\bigcup_{\beta\ge0}\Big\{\beta\Tv_{+}^{(\Kc)}-\Qsf_{\mathrm{inv}}(\Vv^{(\Kc)},\epsilon)\Big\}\bigg\},\label{eq:MD_Local1-12} 
\end{align}
where the first (respectively, second) set is understood to be empty
when $R_{1}^{*}=0$ (respectively, $R_{2}^{*}=0$).
\end{thm}
\begin{IEEEproof}
    See Section \ref{sub:MD_LOCAL_PROOF}.
\end{IEEEproof}
\noindent We make the following remarks on Theorem \ref{thm:MD_LocalResult}:
\begin{enumerate}
    \item In the case that $\hat{\Tv}_{-}=-\hat{\Tv}_{+}$,
    or equivalently $D_{-}=D_{+}$ (i.e.~a curved or straight-line part
    of the boundary), the two sets in \eqref{eq:MD_Local1-12} can be
    combined into a single set containing a coefficient $\beta\in\RR$
    (with negative values allowed) and the vector $\Tv_{+}^{(\Kc)}$.
    In this case, the inner bound on $\Lc(\epsilon,R_{1}^{*},R_{2}^{*})$
    is a half-space. We will see in Section \ref{sub:MD_EXAMPLE2} that
    this does not always occur, and combinations other than $(D_{-},D_{+})=(0,-1)$
    and $(D_{-},D_{+})=(-1,-\infty)$ are possible (these are the combinations
    that are observed for standard pentagonal regions).
    \item In the case that $\Kc$ contains only a single entry $\nu\in\{1,2,12\}$,
    the unions over $\beta$ can be replaced by $\beta=0$, yielding a
    simpler inner bound given by
    \begin{equation}
        \Lc(\epsilon,R_{1}^{*},R_{2}^{*})\supseteq\Big\{(L_{1},L_{2})\,:\, L_{\nu}\le-\sqrt{V_{\nu}}\Qsf^{-1}(\epsilon)\Big\},
    \end{equation}
    where $L_{12}\defeq L_{1}+L_{2}$. The fact that $\beta=0$ suffices
    is shown in the same way for each $\nu$, so we consider the case
    $\nu=12$. Since $(R_{1}^{*},R_{2}^{*})$ lies on the diagonal part
    of the pentagonal region corresponding to $(Q_{U},Q_{1},Q_{2})$ (and
    away from the corners), both $(R_{1}^{*}-\delta,R_{2}^{*}+\delta)$
    and $(R_{1}^{*}+\delta,R_{2}^{*}-\delta)$ are achievable for sufficiently
    small $\delta$, and hence $D_{-}\in[-\infty,-1]$ and $D_{+}\in[-1,0]$.
    The convexity of the capacity region implies that $D_{+}\le D_{-}$,
    and it follows that $D_{-}=D_{+}=-1$. From \eqref{eq:MD_VecsT},
    we see that $D_{-}=D_{+}=-1$ implies $\Tv_{-}^{(\Kc)}=\Tv_{+}^{(\Kc)}=0$,
    and hence each coefficient $\beta$ in \eqref{eq:MD_Local1-12} is
     multiplying zero.
    \item More generally, if the input distribution achieving $(R_{1}^{*},R_{2}^{*})$
    also achieves all of the boundary points in a neighborhood of $(R_{1}^{*},R_{2}^{*})$,
    then the unions over $\beta$ can be replaced by $\beta=0$. In particular,
    this is true when the entire capacity region is achieved by a single
    input distribution. This will be observed for the Gaussian MAC in
    Section \ref{sub:MD_GAUSSIAN}.
    \item All non-empty subsets $\Kc$ of $\{1,2,12\}$ can occur with
    the exception of $\{1,2\}$. Focusing on the case that time-sharing
    is absent, the case $\Kc=\{1,2\}$ is impossible since
    \begin{align}
        I(X_{1},X_{2};Y) & =I(X_{1};Y)+I(X_{2};Y|X_{1})                           \\
                         & \le I(X_{1};X_{2},Y)+I(X_{2};Y|X_{1})                  \\
                         & =I(X_{1};Y|X_{2})+I(X_{2};Y|X_{1}),\label{eq:MD_Ineq3} 
    \end{align}
    where \eqref{eq:MD_Ineq3} follows since $X_{1}$ and $X_{2}$ are
    independent. Whenever $\Kc$ includes $\{1,2\}$, we have
    $R_{1}^{*}=I(X_{1};Y|X_{2})$ and $R_{2}^{*}=I(X_{2};Y|X_{1})$, and
    it follows from \eqref{eq:MD_Ineq3} that $R_{1}^{*}+R_{2}^{*}=I(X_{1},X_{2};Y)$.
    Therefore, we have $\mathcal{K=}\{1,2,12\}$, corresponding to a rectangular
    achievable rate region.
    \item The inner bound in \eqref{eq:MD_Local1-12} is of a similar form to
    the set appearing in \cite[Thm.~3]{PaperCMAC} for the Gaussian MAC
    with degraded message sets. The main differences are (i) The left
    and right tangent vectors are treated separately here, since unlike
    in \cite{PaperCMAC}, the two do not have the same slope in general
    (e.g.~see Figure \ref{fig:MD_ExamplesT} and the example in Section
    \ref{sub:MD_EXAMPLE2}); (ii) There are six cases here corresponding
    to the different subsets $\Kc$ of $\{1,2,12\}$ (see the
    previous item), whereas in \cite{PaperCMAC} there are only two possibilities
    for the set of active rate conditions. 
    \item The proof of Theorem \ref{thm:MD_LocalResult} can be followed using
    i.i.d.~codeword distributions, yielding an analogous result with $\Vv^{\mathrm{iid}}$
    (see \eqref{eq:MD_V_iid}) in place of $\Vv$. Using the
    fact that $\Vv\preceq\Vv^{\mathrm{iid}}$, it
    is not difficult to show that the inner bounds on $\Lc(\epsilon,R_{1}^{*},R_{2}^{*})$
    obtained using $\Vv$ include those obtained using $\Vv^{\mathrm{iid}}$
    whenever $\epsilon<\frac{1}{2}$. In Section \ref{sub:MD_EXAMPLE1},
    we will see that the inclusion can be strict.
    \item It is also of interest to compare $\Vv$ to a hypothetical
    dispersion matrix of the form 
    \begin{equation}
        \Vv^{\mathrm{joint}}\defeq\EE\Big[\cov\big[\iv(U,X_{1},X_{2},Y)\,\big|\, U,X_{1},X_{2}\big]\Big]. \label{eq:MD_V_joint}
    \end{equation}
    This is the matrix that would be obtained if the \emph{joint} composition
    of $(\Uv,\Xv_{1},\Xv_{2})$ were fixed, which is impossible 
    in the absence of cooperation between the users. As we show in
    \cite{PaperMAC2_Conf}, we have $\Vv^{\mathrm{joint}}\preceq\Vv$;
    this is proved using the matrix version of the law of total 
    variance, along with the identity $\EE[\Zv\Zv^{T}]\succeq\EE[\Zv]\EE[\Zv^{T}]$.
    \item We make no attempt to present analytical expressions for $\hat{\Tv}_{-}$
    and $\hat{\Tv}_{+}$, but two numerical approaches to their
    computation are presented in Section \ref{sec:MD_EXAMPLE}. In general,
    if the capacity region is characterized numerically, then one can
    easily obtain numerical bounds or approximations for these tangent
    vectors.
\end{enumerate}

\section{\label{sec:MD_EXAMPLE}Examples}

\subsection{\label{sub:MD_EXAMPLE1}The Collision Channel}

We begin with a simple deterministic example that will permit us
to compare i.i.d.~and constant-composition random coding, and to
discuss the role of the off-diagonal entries of the corresponding 
dispersion matrices.

Setting $\Xc_{1}=\Xc_{2}=\{0,1,2\}$ and 
$\Yc=\{(0,0),(0,1),(0,2),(1,0),(2,0),\csf\}$, the channel
is given by 
\begin{equation}
    W(y|x_{1},x_{2})=\begin{cases}
    1 & y=(x_{1},x_{2})\text{ and }\min\{x_{1},x_{2}\}=0\\
    1 & y=\csf\text{ and }\min\{x_{1},x_{2}\}\ne0\\
    0 & \mathrm{otherwise}.
\end{cases}
\end{equation}
In words, if either user transmits the zero symbol then the pair $(x_{1},x_{2})$
is received noiselessly, whereas if both users transmit a non-zero
symbol then the output is $\csf$, meaning ``collision''. 

\begin{figure}
    \begin{centering}
        \includegraphics[width=0.95\columnwidth]{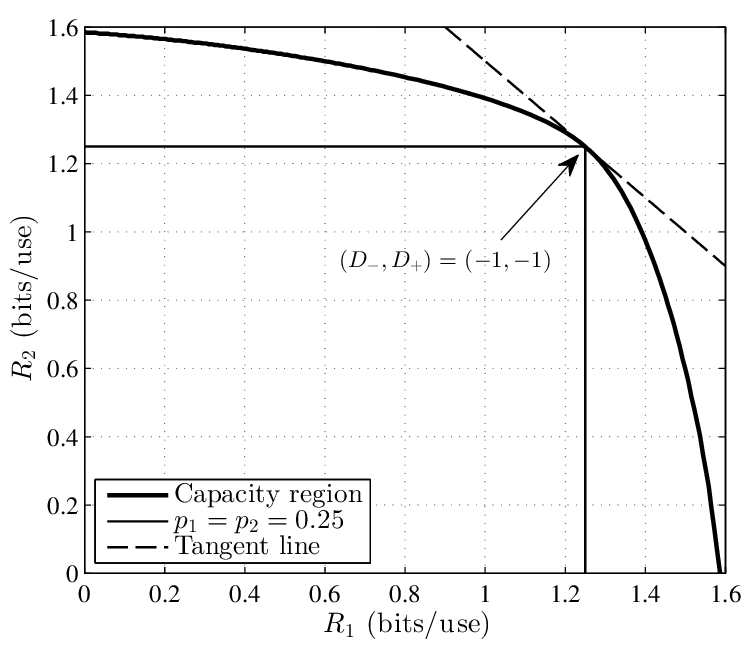}
        \par
    \end{centering}
    
    \protect\caption{Capacity region of the collision channel. The tangent line corresponds
        to the boundary point $(R_{1}^{*},R_{2}^{*})=(1.25,1.25)$, and we
        have $D_{-}=D_{+}=-1$.\label{fig:MD_CapRegion1}}
\end{figure}

We recall the following observations by Gallager \cite{MACExponent2}:
(i) The capacity region can be obtained without time sharing;%
\footnote{On the other hand, for the collision channel with $K$ non-zero symbols,
    time-sharing is required for $K\ge8$ \cite{MACExponent2}.%
} (ii) By symmetry, the points on the boundary of the capacity region
are achieved by input distributions of the form $Q_{1}=(1-2p_{1},p_{1},p_{1})$
and $Q_{2}=(1-2p_{2},p_{2},p_{2})$, where $\Uc=\emptyset$;
(iii) The achievable rate region corresponding to any such $(Q_{1},Q_{2})$
pair is rectangular. The capacity region is shown in Figure \ref{fig:MD_CapRegion1}.
The left and right tangent vectors coincide (i.e.~$\hat{\Tv}_{-}=-\hat{\Tv}_{+}$
and hence $D_{-}=D_{+} \triangleq D$ ) at all boundary points $(R_{1}^{*},R_{2}^{*})$
with $R_{1}^{*}>0$ and $R_{2}^{*}>0$, and the case of interest in
Theorem \ref{thm:MD_LocalResult} is $\Kc=\{1,2,12\}$.

One approach to computing the inner bound in \eqref{eq:MD_Local1-12}
for a given boundary point $(R_{1}^{*},R_{2}^{*})$ is to first find
the pair $(p_{1},p_{2})$ achieving that point, and then calculate
$\hat{\Tv}_{-}$ and $\hat{\Tv}_{+}$ (e.g.
see the example in Section \ref{sub:MD_EXAMPLE2}). In this example,
the reverse approach turns out to be more convenient: We start with
a given derivative $D<0$, and perform an optimization over $(p_{1},p_{2})$
to find the corresponding (unique) boundary point $(R_{1}^{*},R_{2}^{*})$.

As stated above, the achievable rate region for a given pair $(p_{1},p_{2})$
is a rectangle with a corner point given by $(I_{1},I_{2})$. The
straight line of slope $D$ passing through this point is given by
$R_{2}=D(R_{1}-I_{1})+I_{2}$. Thus, finding the desired boundary
point $(R_{1}^{*},R_{2}^{*})$ simply amounts to maximizing $I_{2}-DI_{1}$
with respect to $(p_{1},p_{2})$, which is a straightforward optimization
problem.

\begin{figure}
    \begin{centering}
        \includegraphics[width=0.95\columnwidth]{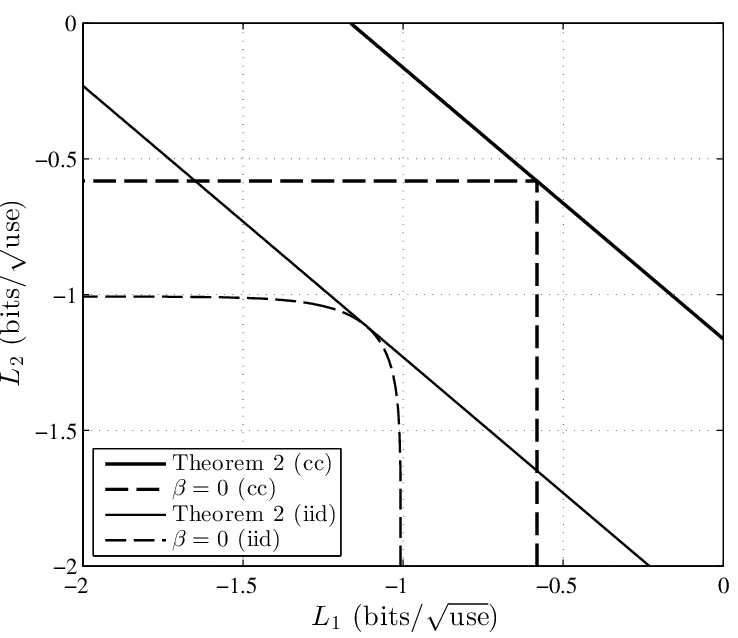}
        \par
    \end{centering}
    
    \protect\caption{Boundaries of the inner bounds on $\Lc(\epsilon,R_{1}^{*},R_{2}^{*})$
        for the collision channel, with $R_{1}^{*}=R_{2}^{*}=1.25$ and $\epsilon=0.01$.
        The regions are to the bottom-left of the boundaries shown. \label{fig:MD_SecondOrderRegion1}}
\end{figure}

For concreteness, we provide a numerical example for the case that
$D=-1$, which corresponds to $\hat{\Tv}_{-}=-\hat{\Tv}_{+}=\big[\frac{-1}{\sqrt{2}}\,\,\,\frac{1}{\sqrt{2}}\big]^{T}$.
Using a brute force search to three decimal places, we found the optimal
parameters to be $p_{1}=p_{2}=0.25$, yielding $R_{1}^{*}=R_{2}^{*}=1.25$
bits/use. The inner bound on $\Lc(\epsilon,R_{1}^{*},R_{2}^{*})$
from Theorem \ref{thm:MD_LocalResult}, and its counterpart for i.i.d.
random coding (cf. \eqref{eq:MD_V_iid}), are shown in Figure \ref{fig:MD_SecondOrderRegion1},
where we set $\epsilon=0.01$. For comparison, we also plot the weaker
inner bounds in which $\beta$ is set to zero and only $-\Qsf_{\mathrm{inv}}$
remains (cf. \eqref{eq:MD_Local1-12}). The boundaries of these regions
are shown, and the regions lie to the bottom-left of these boundaries.

We see that the region resulting from constant-composition random
coding is strictly larger than that resulting from i.i.d.~random coding.
In this example, the strict inclusion holds for all points on the
boundary of the capacity region other than the endpoints corresponding
to $R_{1}^{*}=0$ or $R_{2}^{*}=0$.   This gain
 is analogous to a similar gain in the error exponent
\cite{MACExponent4}. In contrast, in the single-user
setting, the two ensembles yield the same second-order term
and error exponent after the optimization of the input distribution
\cite{Finite,GallagerCC}.

We conclude by discussing the roles of the various
entries of the covariance matrices.  In this example, the
diagonal entries $V_1$ and $V_2$ determine the locations of the
vertical and horizontal asymptotes in Figure 
\ref{fig:MD_SecondOrderRegion1}.  We see
from Figure \ref{fig:MD_SecondOrderRegion1} that the off-diagonal
terms also play a role.  In particular, the rectangular shape of
the region with $\beta=0$ for constant-composition coding is
an extreme case corresponding to a singular dispersion matrix $\Vv$, and
this is in fact the most favorable shape possible (in terms of
enlarging $\Lc(\epsilon;R_1^*,R_2^*)$) given fixed
locations of the vertical and horizontal asymptotes.  In contrast,
the off-diagonal terms of $\Vv^{\mathrm{iid}}$ yield a more
standard curved region for $\beta=0$, which is less favorable.
Thus, at least in this example, the enlarged second-order
region for constant-composition codes is not only due to
smaller diagonal entries, but also due to a more favorable
covariance matrix structure.

\subsection{A Non-Deterministic Example} \label{sub:MD_EXAMPLE2}

Here we provide an example showing that the two unions in \eqref{eq:MD_Local1-12}
cannot, in general, be combined into one. In other words, it is necessary
to consider the left and right tangent vectors separately. We set
$\Xc_{1}=\Xc_{2}=\Yc=\{0,1\}$, and
\begin{equation}
    W(y|x_{1},x_{2})=\begin{cases}
    1 & x_{1}=x_{2}=y\\
    0.5 & x_{1}\ne x_{2}\\
    0 & \mathrm{otherwise}.
\end{cases}\label{eq:MD_Example2}
\end{equation}
Thus, the channel is noiseless if $x_{1}=x_{2}$, and completely noisy
if $x_{1}\ne x_{2}$. We write the input distributions as $Q_{1}=(1-p_{1},p_{1})$
and $Q_{2}=(1-p_{2},p_{2})$.

\begin{figure}
        \begin{centering}
            \includegraphics[width=0.95\columnwidth]{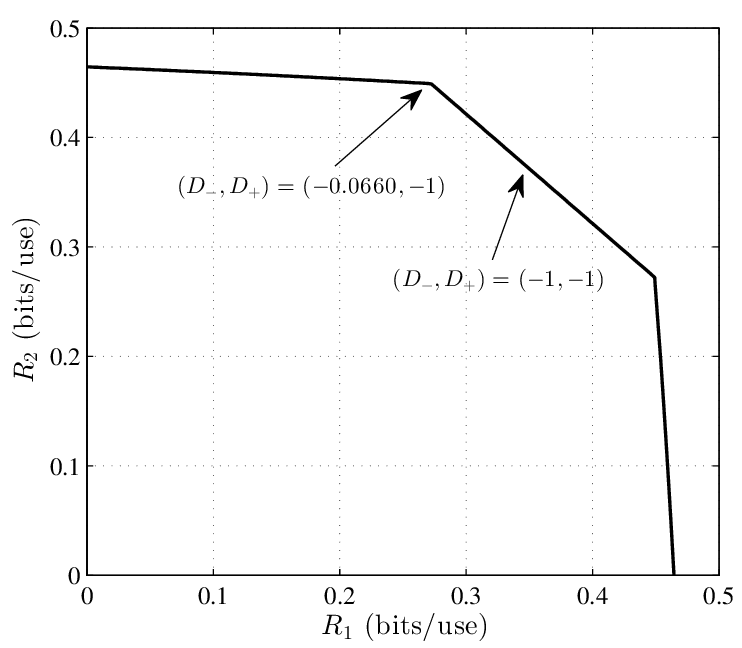}
            \par
        \end{centering} 
        
        \protect\caption{Capacity region of the channel described by \eqref{eq:MD_Example2},
                         and example values of $(D_-,D_+)$.\label{fig:MD_CapRegion2} }
    \end{figure}

The capacity region is shown in Figure \ref{fig:MD_CapRegion2}. Observe
that there are two ``corner points'', but unlike those of standard
pentagonal regions, neither of them corresponds to a change in angle
of 45 degrees. More precisely, the middle segment shown in the plot
has slope $-1$, but the other two segments are neither vertical nor
horizontal (in fact, they are not even straight line segments, even
though they may appear to be). Both corner points are achieved by
$p_{1}=p_{2}=0.5$.

Here we focus on characterizing the set $\Lc(\epsilon,R_{1}^{*},R_{2}^{*})$
for the upper corner point $(R_{1}^{*},R_{2}^{*})=(0.272,0.449)$;
identical arguments apply for the lower corner point. The case of
interest in Theorem \ref{thm:MD_LocalResult} is $\Kc=\{2,12\}$.
Since the middle segment in Figure \ref{fig:MD_CapRegion2} has slope
$-1$, we have $\hat{\Tv}_{+}=\big[\frac{1}{\sqrt{2}}\,\,\frac{-1}{\sqrt{2}}\big]$.
The idea used to compute $\hat{\Tv}_{-}$ is to shift the
point $(p_{1},p_{2})=(0.5,0.5)$ by a small amount $(\Delta\cos\theta,\Delta\sin\theta)$,
and observe the behavior of the corner point $(I_{12}-I_{2},I_{2})$
for $\theta\in[0,2\pi)$, where for each $\theta$ we are interested
    in the limiting behavior as $\Delta\to0$. Making the dependence of
    $I_{2}$ and $I_{12}$ on $(p_{1},p_{2})$ explicit, a second-order
    Taylor expansion yields
    \begin{equation}
        I_{\nu}(p_{1}+\Delta\cos\theta,p_{2}+\Delta\sin\theta)=I_{\nu}(p_{1},p_{2})+f_{\nu}(\theta)\Delta^{2}+O(\Delta^{3}),\label{eq:MD_ExampleTaylor}
    \end{equation}
    for $\nu=2,12$, where
    \begin{equation}
        f_{\nu}(\theta)\defeq\big[\cos\theta\,\,\sin\theta\big]\left[\begin{array}{cc}
            \frac{\partial^{2}I_{\nu}}{\partial p_{1}^{2}} & \frac{\partial^{2}I_{\nu}}{\partial p_{1}\partial p_{2}}\\
            \frac{\partial^{2}I_{\nu}}{\partial p_{1}\partial p_{2}} & \frac{\partial^{2}I_{\nu}}{\partial p_{2}^{2}}
            
            \end{array}\right]\left[\begin{array}{c}
            \cos\theta\\
            \sin\theta
            
        \end{array}\right].
    \end{equation}
    Note that the first-order term in \eqref{eq:MD_ExampleTaylor} is
    absent, since the derivatives $\frac{\partial I_{\nu}}{\partial p_{1}}$
    and $\frac{\partial I_{\nu}}{\partial p_{2}}$ are zero at $(p_{1},p_{2})=(0.5,0.5)$
    for $\nu=2,12$. We conclude from \eqref{eq:MD_ExampleTaylor} that
    for a fixed choice of $\theta$, $(I_{12}-I_{2},I_{2})$ moves in
    the direction $(f_{12}(\theta)-f_{2}(\theta),f_{2}(\theta))$ in the
    limit as $\Delta\to0$. Evaluating the direction for 10000 equally
    spaced angles over the range $\theta\in[0,2\pi]$, we obtained $\hat{\Tv}_{-}=\big[{-0.99783}\,\,\,0.06586\big]$,
    corresponding to $\theta=0.221$ radians and $D_{-}=-0.0660$.
    
    \begin{figure}
        \begin{centering}
            \includegraphics[width=0.95\columnwidth]{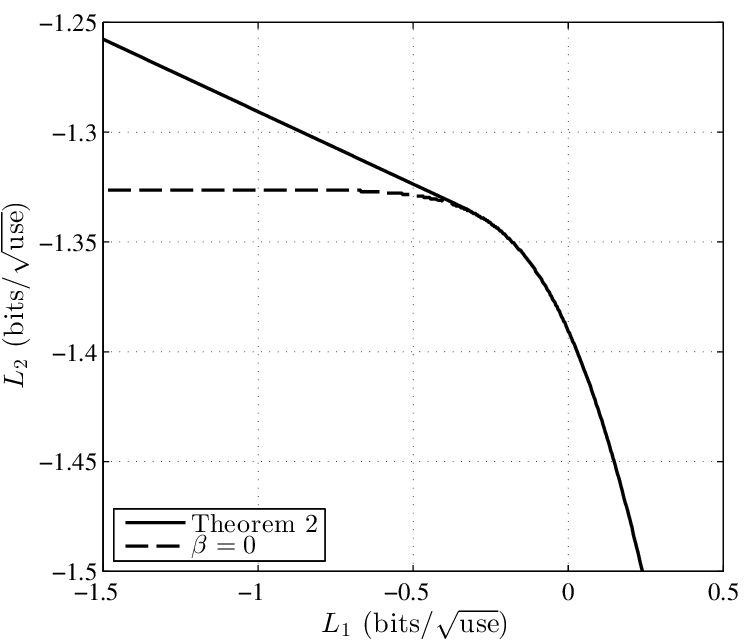}
            \par
        \end{centering}
         
        \protect\caption{Boundaries of the inner bounds on $\Lc(\epsilon,R_{1}^{*},R_{2}^{*})$
            for the channel described by \eqref{eq:MD_Example2}, with $(R_{1}^{*},R_{2}^{*})=(0.272,0.449)$
            and $\epsilon=0.01$. The regions are to the bottom-left of the boundaries
            shown. \label{fig:MD_SecondOrderRegion2}}
    \end{figure}
    
    Figure \ref{fig:MD_SecondOrderRegion2} shows the resulting inner
    bound on $\Lc(\epsilon,R_{1}^{*},R_{2}^{*})$ given in Theorem
    \ref{thm:MD_LocalResult}, with $\epsilon=0.01$. In this example,
    the region is identical for i.i.d.~random coding and constant-composition
    random coding. It is interesting to observe the different shape of
    the region compared to the previous example, resulting from the differing
    left and right tangent vectors. It is only the former that plays a
    role in enlarging the achievable region, since the set $-\Qsf_{\mathrm{inv}}(\Vv^{(\Kc)},\epsilon)$
    (with $\Kc = \{2,12\}$) already satisfies the property that if a given point is in the set,
    so are all points on the right-hand side of the line of slope $D_{+}=-1$
    passing through that point.
    
    \subsection{\label{sub:MD_GAUSSIAN}Gaussian Multiple-Access Channel}
    
    We have focused our attention on the DM-MAC, which permits an analysis
    based on combinatorial arguments. We now discuss how Theorem \ref{thm:MD_LocalResult}
    can be extended to the Gaussian MAC via an increasingly fine quantization
    of the inputs, similarly to Hayashi \cite[Thm.~3]{Hayashi} and Tan
    \cite{TanWiretap}. Each use of the channel is described by
    \begin{equation}
        Y=\sqrt{P_{1}}X_{1}+\sqrt{P_{2}}X_{2}+Z,\label{eq:MD_GaussianChannel}
    \end{equation}
    where $Z\sim N(0,1)$, and where each codeword $\xv_\nu$ for user $\nu=1,2$
    is constrained to satisfy $\frac{1}{n}\|\xv_{\nu}\|^{2}\le1$.
    The quantities $P_{1}$ and $P_{2}$ represent the signal-to-noise
    ratios for users 1 and 2 respectively.
    
    The capacity region is pentagonal \cite[Sec.~15.1]{Cover}, and is
    achieved using Gaussian input distributions, namely $Q_{1},Q_{2}\sim N(0,1)$.
    The quantities $\Iv$ and $\Vv$ in \eqref{eq:MD_VecI}
    and \eqref{eq:MD_CovMatrix} can be written explicitly;
    for $\nu=1,2,12$, we have
    \begin{equation}
        I_{\nu} = \frac{1}{2}\log\big( 1 + P_{\nu} \big) \label{eq:MD_GaussianI}
    \end{equation}
    with $P_{12} \defeq P_1 + P_2$.  Moreover, for $\nu=1,2$,
    we have 
    \begin{align}
        V_{\nu} &= \frac{P_{\nu}(2+P_{\nu})}{2(1+P_{\nu})^{2}} \\
        V_{\nu,12} &= \frac{P_{\nu}(2+P_{1}+P_{2})}{2(1+P_{\nu})(1+P_{1}+P_{2})},
    \end{align}
    and the remaining entries of $\Vv$ are given by
    \begin{align}
        V_{12} &= \frac{(P_{1}+P_{2})(2+P_{1}+P_{2})+2P_{1}P_{2}}{2(1+P_{1}+P_{2})^{2}} \\
        V_{1,2} &= \frac{P_{1}P_{2}}{2(1+P_{1})(1+P_{2})}. \label{eq:MD_GaussianV12}
    \end{align}
    A brief outline of how these expressions are obtained is
    given in Appendix \ref{sub:MD_GaussConv}.
    
    
%
    
    We claim that the following inner bound holds using the notation of
    Theorem \ref{thm:MD_LocalResult} (with the additional condition that
    the codewords must satisfy the above power constraints
    in the definition of $\Lc$ in Section \ref{sub:MD_SYSTEM_MODEL}):
    \begin{equation}
        \Lc(\epsilon,R_{1}^{*},R_{2}^{*})\supseteq\Big\{(L_{1},L_{2})\,:\,\Lv^{(\Kc)}\in-\Qsf_{\mathrm{inv}}(\Vv^{(\Kc)},\epsilon)\Big\}.\label{eq:MD_Gaussian2OR}
    \end{equation}
    This result was first derived by MolavianJazi
    and Laneman \cite{MolavianJazi}, who used random coding according
    to the uniform distribution over the surface of a sphere. The techniques
    of this paper provide an alternative approach to deriving the result.
    Extending Theorem \ref{thm:MD_LocalResult} accordingly is non-trivial,
    but it is done using well-established techniques; we provide
    an outline in Appendix \ref{sub:MD_GaussConv}. In contrast with
    Theorem \ref{thm:MD_LocalResult}, no form of time-sharing is used
    in the proof, and no tangent vectors appear in \eqref{eq:MD_Gaussian2OR}.
    This is due to the fact that every point on the boundary of the capacity
    region is simultaneously achieved by Gaussian inputs.

    \section{\label{sec:MD_PROOF}Proof of Theorem \ref{thm:MD_LocalResult}}
    
    The random-coding ensemble used in the proof depends on 
    two triplets $(Q_{U},Q_{1},Q_{2})$ and $(Q_{U}^{\prime},Q_{1}^{\prime},Q_{2}^{\prime})$
    of probability distributions on the same alphabets. We define $P_{UX_{1}X_{2}Y}^{\prime}$, $\iv^{\prime}$,
    $\iv^{\prime(1)}$, $\iv^{\prime(2)}$, $\Iv^{\prime}$
    and $\Vv^{\prime}$ in the same way as Sections \ref{sub:MD_SYSTEM_MODEL}
    and \ref{sub:MD_PRELIM}, with $(Q_{U}^{\prime},Q_{1}^{\prime},Q_{2}^{\prime})$
    replacing $(Q_{U},Q_{1},Q_{2})$. In particular, we have
    \begin{align}
        \Iv^{\prime} & \defeq\EE\big[\iv^{\prime}(U^{\prime},X_{1}^{\prime},X_{2}^{\prime},Y^{\prime})\big]\label{eq:MD_VectorI'}                                                                                                                                                                                                                                              \\
        \Vv^{\prime} & \defeq\EE\Big[\cov\big[\iv^{\prime}(U^{\prime},X_{1}^{\prime},X_{2}^{\prime},Y^{\prime})\,\big|\, U^{\prime}\big] \nonumber \\ 
            & \quad -\cov\big[\iv^{\prime(1)}(U^{\prime},X_{1}^{\prime})\,\big|\, U^{\prime}\big]-\cov\big[\iv^{\prime(2)}(U^{\prime},X_{2}^{\prime})\,\big|\, U^{\prime}\big]\Big],\label{eq:MD_CovMatrix'} 
    \end{align}
    where $(U^{\prime},X_{1}^{\prime},X_{2}^{\prime},Y^{\prime})\sim P_{UX_{1}X_{2}Y}^{\prime}$.
    As an intermediate step towards obtaining our local result, we present
    the following global result.
    \begin{thm}
        \label{thm:MD_MainResult} Fix any finite time-sharing alphabet $\Uc$
        and the input distributions $(Q_{U},Q_{1},Q_{2})$ and $(Q_{U}^{\prime},Q_{1}^{\prime},Q_{2}^{\prime})$.
        For any $\beta\ge0$ and $\epsilon\in(0,1)$, there exists a function
        $g(n)=O\big(n^{\frac{1}{4}}\big)$ (depending on $(Q_{U},Q_{1},Q_{2})$,
        $(Q_{U}^{\prime},Q_{1}^{\prime},Q_{2}^{\prime})$, $\beta$ and $\epsilon$)
        such that all rate pairs $(R_{1},R_{2})$ satisfying 
        \begin{equation}
            n\Rv\in n\Iv+\sqrt{n}\Big(\beta(\Iv^{\prime}-\Iv)-\Qsf_{\mathrm{inv}}(\Vv,\epsilon)\Big)+g(n)\bone\label{eq:MD_Achievable-2}            
        \end{equation}
        are $(n,\epsilon)$-achievable.
    \end{thm}
    \begin{IEEEproof}
        For clarity of exposition, we present the proof in two parts. In Section
        \ref{sub:MD_PROOF_SIMPLIFIED}, we handle the case that $\beta=0$
        and $\Uc=\emptyset$. The changes required to handle the general
        case are presented in Section  \ref{sub:MD_PROOF_PART2}.
    \end{IEEEproof}
    In the case that $\beta=0$, Theorem \ref{thm:MD_MainResult} gives
    the second-order asymptotics for a fixed triplet $(Q_{U},Q_{1},Q_{2})$.
    In this case, the proof reveals that the behavior $g(n)=O\big(n^{\frac{1}{4}}\big)$
    can be strengthened to $g(n)=O(\log n)$ if the argument to the expectation
    in \eqref{eq:MD_CovMatrix} has full rank for all $u$, and $g(n)=O\big(n^{\frac{1}{6}}\big)$
    more generally.
    
    The case $\beta>0$ is proved by applying an extended version of coded
    time-sharing \cite[Sec.~4.5.3]{NetworkBook} in which a fraction $1-\frac{\beta}{\sqrt{n}}$
    of the symbols are generated using $(Q_{U},Q_{1},Q_{2})$, and the
    remaining symbols are generated using $(Q_{U}^{\prime},Q_{1}^{\prime},Q_{2}^{\prime})$.

    \subsection{\label{sub:MD_LOCAL_PROOF}Proof of Theorem \ref{thm:MD_LocalResult}
        Based on Theorem \ref{thm:MD_MainResult}}
    
    Throughout the proof, we write $R_{12}\defeq R_{1}+R_{2}$ and
    $R_{12}^{*}\defeq R_{1}^{*}+R_{2}^{*}$ (recall that $(R_{1}^{*},R_{2}^{*})$
    is the boundary point of interest). We have from the definition of
    $\Kc$ that $R_{\nu}^{*}<I_{\nu}$ for all $\nu\in\Kc^{c}$,
    and we can thus weaken \eqref{eq:MD_Achievable-2} to
    \begin{equation}
        n\Rv\in n\Rv^{*}+n\delta\bone_{\Kc^{c}}+\sqrt{n}\Big(\beta(\Iv^{\prime}-\Iv)-\Qsf_{\mathrm{inv}}(\Vv,\epsilon)\Big)+O\big(n^{\frac{1}{4}}\big)\bone,\label{eq:MD_LocalStep4}
    \end{equation}
    where $\delta\defeq\min_{\nu\in\Kc^{c}}I_{\nu}-R_{\nu}^{*}>0$,
    and $\bone_{\Kc^{c}}$ contains ones at the indices
    corresponding to $\Kc^{c}$, and zeros elsewhere. 
    
    Let $\zv$ be a $3\times1$ vector, and let $\zv^{(\Kc)}$
    and $\zv^{(\Kc^{c})}$ be the corresponding subvectors
    indexed by the superscript. From the definition of $\Qsf_{\mathrm{inv}}$,
    the set $-\Qsf_{\mathrm{inv}}(\Vv^{(\Kc)},\epsilon)$
    contains the vectors $\zv^{(\Kc)}$  such that
    $\zv\in-\Qsf_{\mathrm{inv}}(\Vv,\epsilon)$
    in the limit as the entries of $\zv^{(\Kc^{c})}$
    tend towards $-\infty$ \cite{NomuraHan}. Moreover, since $n\delta$
    grows faster than $O(\sqrt{n})$, the elements of $n\Rv$
    corresponding to $\Kc^{c}$ in \eqref{eq:MD_LocalStep4} may
    incur an additional $L\sqrt{n}$ term for \emph{any} value of $L$.
    Combining these observations with the definition of $\Lc$,
    we obtain
    \begin{multline}
        \Lc(\epsilon,R_{1}^{*},R_{2}^{*})\supseteq\bigg\{(L_{1},L_{2})\,:\, \\ \Lv^{(\Kc)}\in\bigcup_{\beta\ge0}\Big\{\beta(\Iv^{\prime(\Kc)}-\Iv^{(\Kc)})-\Qsf_{\mathrm{inv}}(\Vv^{(\Kc)},\epsilon)\Big\}\bigg\}.\label{eq:MD_LocalStep6}
    \end{multline}
    Suppose for the time being that $R_{2}^{*}>0$, and let $(Q_{U}^{\prime},Q_{1}^{\prime},Q_{2}^{\prime})$
    be chosen to achieve a boundary point $(R_{1}^{\prime},R_{2}^{\prime})$
    to the right of $(R_{1}^{*},R_{2}^{*})$ (more precisely, one such
    that $R_{1}^{\prime}\ge R_{1}^{*}$ and $R_{2}^{\prime}\le R_{2}^{*}$,
    with at least one of the inequalities being strict), and define $R_{12}^{\prime}$,
    $\Rv^{\prime}$ and $\Rv^{\prime(\Kc)}$
    in the same way as $R_{12}$, $\Rv$ and $\Rv^{(\Kc)}$.
    Since $(R_{1}^{\prime},R_{2}^{\prime})$ is achieved by $(Q_{U}^{\prime},Q_{1}^{\prime},Q_{2}^{\prime})$,
    we have $\Rv^{\prime}\preceq\Iv^{\prime}$,
    and in particular, $\Rv^{\prime(\Kc)}\preceq\Iv^{\prime(\Kc)}$.
    Moreover, from the definition of $\Kc$, we have $\Iv^{(\Kc)}=\Rv^{*(\Kc)}$.
    Combining these, we deduce from \eqref{eq:MD_LocalStep6} that
    \begin{multline}
        \Lc(\epsilon,R_{1}^{*},R_{2}^{*})\supseteq\bigg\{(L_{1},L_{2})\,:\, \\ \Lv^{(\Kc)}\in\bigcup_{\beta\ge0}\Big\{\beta(\Rv^{\prime(\Kc)}-\Rv^{*(\Kc)})-\Qsf_{\mathrm{inv}}(\Vv^{(\Kc)},\epsilon)\Big\}\bigg\}.\label{eq:MD_LocalStep7}
    \end{multline}
    By the definition of $\Tv_{+}$ (see \eqref{eq:MD_VecsT}),
    the direction%
    \footnote{If $|\Kc|=1$, the ``direction'' should be interpreted as
        being the sign.%
        } of the vector $\Rv^{\prime(\Kc)}-\Rv^{*(\Kc)}$
approaches that of $\Tv_{+}^{(\Kc)}$ as $(R_{1}^{\prime},R_{2}^{\prime})$
approaches $(R_{1}^{*},R_{2}^{*})$ along the boundary from the right.
By taking this limiting choice and using the fact that $\Lc$
is defined using a closure operation, we obtain the second set in
\eqref{eq:MD_Local1-12}. Provided that $R_{1}^{*}>0$, the first
set is obtained in an identical fashion by letting $(Q_{U}^{\prime},Q_{1}^{\prime},Q_{2}^{\prime})$
achieve a boundary point approaching $(R_{1}^{*},R_{2}^{*})$ from
the left.

\subsection{\label{sub:MD_PROOF_SIMPLIFIED}Proof of Theorem \ref{thm:MD_MainResult}
    ($\beta=0$, $\Uc=\emptyset$)}

In this subsection, we consider the case that $\beta=0$ and $\Uc=\emptyset$,
and we omit the arguments $u$ to the functions defined in Section
\ref{sub:MD_PRELIM} (e.g.~$\iv(u,x_{1},x_{2},y)$ is replaced
by $\iv(x_{1},x_{2},y)$). 

For $\nu=1,2$, we are given the input distribution $Q_{\nu}\in\Pc(\Xc_{\nu})$,
and we let $Q_{\nu,n}\in\Pc_{n}(\Xc_{\nu})$ be a
type with the same support as $Q_{\nu}$ such that $\max_{x_{\nu}}|Q_{\nu}(x_{\nu})-Q_{\nu,n}(x_{\nu})|\le\frac{1}{n}$.
We generate the $M_{\nu}\defeq e^{nR_{\nu}}$ codewords of user
$\nu=1,2$ independently according to the uniform distribution on
$T^{n}(Q_{\nu,n})$, i.e.~
\begin{equation}
    P_{\Xv_{\nu}}(\xv_{\nu})=\frac{1}{|T^{n}(Q_{\nu,n})|}\openone\Big\{\xv_{\nu}\in T^{n}(Q_{\nu,n})\Big\}.\label{eq:MD_Pxv}
\end{equation}
For clarity of exposition, we assume that $Q_{1}$ and $Q_{2}$ are
themselves types, and hence $Q_{1,n}=Q_{1}$ and $Q_{2,n}=Q_{2}$;
the analysis for the more general case introduces an additive $O(1)$
term that can be incorporated into $g(n)$ in \eqref{eq:MD_Achievable-2}.

We define the random variables 
\begin{multline} 
    (\Xv_{1},\Xv_{2},\Yv,\overline{\Xv}_{1},\overline{\Xv}_{2})\sim P_{\Xv_{1}}(\xv_{1})P_{\Xv_{2}}(\xv_{2})W^{n}(\yv|\xv_{1},\xv_{2}) \\ \times P_{\Xv_{1}}(\overline{\xv}_{1})P_{\Xv_{2}}(\overline{\xv}_{2}).\label{eq:MD_Distribution}
\end{multline} 
Using a threshold-based decoder and standard bounding techniques (e.g.
see \cite{MolavianJazi}), we can upper bound the random-coding
error probability $\overline{p}_{e}$ as follows:
\begin{align}
    \overline{p}_{e} & \le1-\PP\Big[\iv^{n}(\Xv_{1},\Xv_{2},\Yv)\succ\bgamma\Big] \nonumber \\
                     & \quad+M_{1}\PP\Big[i_{1}^{n}(\overline{\Xv}_{1},\Xv_{2},\Yv)>\gamma_{1}\Big]\nonumber                           \\
                     & \quad+M_{2}\PP\Big[i_{2}^{n}(\Xv_{1},\overline{\Xv}_{2},\Yv)>\gamma_{2}\Big] \nonumber \\
                     & \quad+M_{1}M_{2}\PP\Big[i_{12}^{n}(\overline{\Xv}_{1},\overline{\Xv}_{2},\Yv)>\gamma_{12}\Big],\label{eq:MD_pe_bar} 
\end{align}
where $\bgamma=[\gamma_{1}\:\gamma_{2}\:\gamma_{12}]^{T}$
is arbitrary, and 
\begin{align}
    \iv^{n}(\xv_{1},\xv_{2},\yv) & \defeq\sum_{i=1}^{n}\iv(x_{1,i},x_{2,i},y_{i})\label{eq:MD_VectorInBold} \\
    i_{\nu}^{n}(\xv_{1},\xv_{2},\yv)        & \defeq\sum_{i=1}^{n}i_{\nu}(x_{1,i},x_{2,i},y_{i}).\label{eq:MD_VectorIn}           
\end{align}
By applying a standard change of measure from constant-composition
to i.i.d.~\cite{CsiszarBook,Hayashi}, we can upper bound the second,
third and fourth terms of \eqref{eq:MD_pe_bar} by $M_{\nu}p_{0}(n)e^{-\gamma_{\nu}}$
for $\nu=1,2,12$ respectively, where $p_{0}(n)\defeq(n+1)^{|\Xc_{1}|+|\Xc_{2}|-2}$,
and $M_{12}\defeq M_{1}M_{2}.$ We thus obtain
\begin{align}
    \overline{p}_{e} & \le1-\PP\big[\iv^{n}(\Xv_{1},\Xv_{2},\Yv)\succ\bgamma\big]+p_{0}(n)\sum_{\nu=1,2,12}M_{\nu}e^{-\gamma_{\nu}}.\label{eq:MD_pe_bar2} 
\end{align}
Using this bound with
\begin{equation}
    \gamma_{\nu}=\log M_{\nu}+\Big(d+\frac{1}{2}\Big)\log n,\label{eq:MD_gamma}
\end{equation}
where $d\defeq|\Xc_{1}|+|\Xc_{2}|-2$ is the order
of the polynomial $p_{0}(n)$, the desired result in \eqref{eq:MD_Achievable-2}
(with $\beta=0$ and $\Uc=\emptyset$) will follow using nearly identical
steps to \cite[Thm.~4]{MACFinite1} once we prove the following:
\begin{enumerate}
    \item $\EE[\iv^{n}(\Xv_{1},\Xv_{2},\Yv)]=n\Iv$
    and $\cov[\iv^{n}(\Xv_{1},\Xv_{2},\Yv)]=n\Vv+\bDelta_n$,
    where $\Iv$ and $\Vv$ are given in \eqref{eq:MD_VecI}
    and \eqref{eq:MD_CovMatrix}, and $\bDelta_n$ has $O(1)$ entries. 
    \item In the case that $\Vv\succ\bzero$, the probability
    on the right-hand side of \eqref{eq:MD_pe_bar2} can be approximated
    using a multivariate Berry-Esseen theorem with $O\big(\frac{1}{\sqrt{n}}\big)$
    convergence.  
    \item In the case that $\Vv$ is singular, the problem can be
    reduced to a lower dimension using Chebyshev's inequality,
    and the Berry-Esseen theorem can again be applied.
\end{enumerate}
We formalize and prove these statements in the remainder of this subsection;
the remaining details of the proof of \eqref{eq:MD_Achievable-2}
are omitted to avoid repetition with \cite{MACFinite1}.

\subsubsection{\label{sub:MD_MOMENTS}Calculation of Moments}

The first moment of $\iv^{n}$ is easily found by writing
\begin{equation}
    \EE\big[\iv^{n}(\Xv_{1},\Xv_{2},\Yv)\big]=\sum_{i=1}^{n}\EE\big[\iv(X_{1,i},X_{2,i},Y_{i})\big]=n\Iv,
\end{equation}
where the last equality follows since, by symmetry, $X_{1,i}\sim Q_{1}$
and $X_{2,i}\sim Q_{2}$ for all $i$. To compute the covariance matrix
of $\iv^{n}$, we write 
\begin{align}
    &\cov\Big[\iv^{n}(\Xv_{1},\Xv_{2},\Yv)\Big] \nonumber \\
    & \quad=\cov\bigg[\sum_{i=1}^{n}\iv(X_{1,i},X_{2,i},Y_{i})\bigg]                                                                                                                          \\
    & \quad=\sum_{i=1}^{n}\sum_{j=1}^{n}\cov\Big[\iv(X_{1,i},X_{2,i},Y_{i}),\iv(X_{1,j},X_{2,j},Y_{j})\Big]\label{eq:MD_VarStep2} \\
    & \quad=n\cov\Big[\iv(X_{1},X_{2},Y)\Big] \nonumber \\
    & \quad\quad +(n^{2}-n)\cov\Big[\iv(X_{1},X_{2},Y),\iv(X_{1}^{\dagger},X_{2}^{\dagger},Y^{\dagger})\Big],\label{eq:MD_VarStep3} 
\end{align}
where $(X_{1},X_{2},Y)$ and $(X_{1}^{\dagger},X_{2}^{\dagger},Y^{\dagger})$
correspond to two arbitrary but different indices in $\{1,\cdots,n\}$
(e.g.~one can set $(X_{1},X_{2},Y)=(X_{1,1},X_{2,1},Y_{1})$ and $(X_{1}^{\dagger},X_{2}^{\dagger},Y^{\dagger})=(X_{1,2},X_{2,2},Y_{2})$).
Equation \eqref{eq:MD_VarStep3} follows by noting that the symmetry of
the codebook construction implies that
the $n$ terms in \eqref{eq:MD_VarStep2} with $i=j$ are equal, 
and similarly for the $n^{2}-n$ terms with $i\ne j$.

To compute the cross-covariance matrix in \eqref{eq:MD_VarStep3},
we need the joint distribution of $(X_{1},X_{2},Y)$ and $(X_{1}^{\dagger},X_{2}^{\dagger},Y^{\dagger})$.
This distribution can be understood by noting that the uniform distribution
on $T^{n}(Q)$ is obtained by randomly permuting the symbols of an
arbitrary sequence $\xv\in T^{n}(Q)$. This, in turn, can
be interpreted as successively performing uniform sampling from a
collection of symbols without replacement ($n$ times in total), where
the initial collection contains $nQ(x)$ occurrences of each symbol
$x\in\Xc$. By considering the first two steps of such a procedure,
we readily obtain
\begin{align}
    \PP[X_{\nu}=x_{\nu}]                                     & =Q_{\nu}(x_{\nu})\label{eq:MD_SingleProb}                                                            \\
    \PP[X_{\nu}^{\dagger}=x_{\nu}^{\dagger}|X_{\nu}=x_{\nu}] & =\frac{nQ_{\nu}(x_{\nu}^{\dagger})-\openone\{x_{\nu}=x_{\nu}^{\dagger}\}}{n-1}\label{eq:MD_CondProb} 
\end{align}
for $\nu=1,2$. Letting $P_{\nu}^{\dagger}(x_{\nu}^{\dagger}|x_{\nu})$
denote the right-hand side of \eqref{eq:MD_CondProb}, the cross-covariance
matrix in \eqref{eq:MD_VarStep3} is given by 
\begin{align}
      & \cov\Big[\iv(X_{1},X_{2},Y),\iv(X_{1}^{\dagger},X_{2}^{\dagger},Y^{\dagger})\Big]\nonumber                                                                                                                                                                                                                          \\
      & \quad=\EE\Big[\big(\iv(X_{1},X_{2},Y)-\Iv\big)\big(\iv(X_{1}^{\dagger},X_{2}^{\dagger},Y^{\dagger})-\Iv\big)^{T}\Big]                                                                                                                                                                          \\
      & \quad=\sum_{x_{1},x_{2},y}Q_{1}(x_{1})Q_{2}(x_{2})W(y|x_{1},x_{2}) \nonumber \\
      & \qquad\times\sum_{x_{1}^{\dagger},x_{2}^{\dagger},y^{\dagger}}P_{1}^{\dagger}(x_{1}^{\dagger}|x_{1})P_{2}^{\dagger}(x_{2}^{\dagger}|x_{2})W(y^{\dagger}|x_{1}^{\dagger},x_{2}^{\dagger}) \nonumber \\
      & \qquad\times\big(\iv(x_{1},x_{2},y)-\Iv\big)\big(\iv(x_{1}^{\dagger},x_{2}^{\dagger},y^{\dagger})-\Iv\big)^{T} \\
      & \quad\defeq\Fv_{1}+\Fv_{2}+\Fv_{3}+\Fv_{4},\label{eq:MD_SumOfTs}                                                                                                                                                                                                                                  
\end{align}
where the four terms in \eqref{eq:MD_SumOfTs} correspond to the four
terms in the expansion of $\big(nQ_{1}(x_{1}^{\dagger})-\openone\{x_{1}=x_{1}^{\dagger}\}\big)\big(nQ_{2}(x_{2}^{\dagger})-\openone\{x_{2}=x_{2}^{\dagger}\}\big)$
resulting from \eqref{eq:MD_CondProb}. These can be written as 
\begin{align}
    \Fv_{1} & =\frac{n^{2}}{(n-1)^{2}}\EE\Big[\iv(X_{1},X_{2},Y)-\Iv\Big]\EE\Big[\iv(X_{1},X_{2},Y)-\Iv\Big]^{T}\label{eq:MD_T1}                    \\
    \Fv_{2} & =-\frac{n}{(n-1)^{2}}\EE\Big[\big(\iv(X_{1},X_{2},Y)-\Iv\big)\big(\iv(\overline{X}_{1},X_{2},\overline{Y})-\Iv\big)^{T}\Big]\label{eq:MD_T2} \\
    \Fv_{3} & =-\frac{n}{(n-1)^{2}}\EE\Big[\big(\iv(X_{1},X_{2},Y)-\Iv\big)\big(\iv(X_{1},\overline{X}_{2},\overline{\overline{Y}})-\Iv\big)^{T}\Big]      \\
    \Fv_{4} & =\frac{1}{(n-1)^{2}}\EE\Big[\big(\iv(X_{1},X_{2},Y)-\Iv\big)\big(\iv(X_{1},X_{2},\Ytilde)-\Iv\big)^{T}\Big],\label{eq:MD_T4}           
\end{align}
where 
\begin{align}
    &(X_{1},X_{2},Y,\overline{X}_{1},\overline{X}_{2},\overline{Y},\overline{\overline{Y}},\Ytilde)\sim Q_{1}(x_{1})Q_{2}(x_{2})W(y|x_{1},x_{2}) \nonumber \\ 
    &~~~~~~\times Q_{1}(\overline{x}_{1})Q_{2}(\overline{x}_{2})W(\overline{y}|\overline{x}_{1},x_{2})W(\overline{\overline{y}}|x_{1},\overline{x}_{2})W(\widetilde{y}|x_{1},x_{2}).
\end{align}
Observe that $\Fv_{1}$ is the zero matrix since $\Iv$ is the mean of $\iv$, and $\Fv_{4}$ 
has $O\big(\frac{1}{n^{2}}\big)$ entries since the expectation does not depend on $n$.
Furthermore, recalling the definition of $\iv^{(2)}$ in
\eqref{eq:MD_ivec_2}, the expectation in \eqref{eq:MD_T2} can be
written as 
\begin{align}
      & \EE\bigg[\EE\Big[\big(\iv(X_{1},X_{2},Y)-\Iv\big)\Big|X_{2}\Big]\EE\Big[\big(\iv(\overline{X}_{1},X_{2},\overline{Y})-\Iv\big)\Big|X_{2}\Big]^{T}\bigg] \\
      & \quad=\cov\big[\iv^{(2)}(X_{2})\big].                                                                                                                                                                 
\end{align}
It follows that
\begin{equation}
    \Fv_{2}=\frac{-n}{(n-1)^{2}}\cov\big[\iv^{(2)}(X_{2})\big],\label{eq:MD_Cov2}
\end{equation}
and we similarly have 
\begin{equation}
    \Fv_{3}=\frac{-n}{(n-1)^{2}}\cov\big[\iv^{(1)}(X_{1})\big].\label{eq:MD_Cov1}
\end{equation}
Using the identity $\frac{n}{(n-1)^{2}}=\frac{1}{n}+O\big(\frac{1}{n^{2}}\big)$
and combining \eqref{eq:MD_VarStep3}, \eqref{eq:MD_SumOfTs}, \eqref{eq:MD_Cov2}
and \eqref{eq:MD_Cov1}, we obtain
\begin{equation}
    \cov\big[\iv^{n}(\Xv_{1},\Xv_{2},\Yv)\big]=n\Vv+\bDelta_n,\label{eq:MD_CovFinal}
\end{equation} 
where $\Vv$ is defined as in \eqref{eq:MD_CovMatrix} with $\Uc=\emptyset$,
and $\bDelta_n$ has $O(1)$ entries.

\subsubsection{\label{sub:MD_BERRY_ESSEEN}A Combinatorial Berry-Esseen Theorem}

The Berry-Esseen theorem required to bound the probability in \eqref{eq:MD_pe_bar2}
is a special case of a result by Loh \cite[Thm.~2]{CombLatin} for
a problem known as Latin hypercube sampling. This result builds on
a combinatorial central limit theorem due to Hoeffding \cite{CombHoeffding};
for other related works, see \cite{CombBolt,CombHoChen,CombMultVar,CombVonBahr}
and the references therein.

We define the quantities 
\begin{align}
    \bSigma_{n}               & \defeq\frac{1}{n}\cov\big[\iv^{n}(\Xv_{1},\Xv_{2},\Yv)\big],\label{eq:MD_SigmaN}                                                               \\
    \widehat{\Sv}_{n}          & \defeq\frac{1}{\sqrt{n}}\bSigma_{n}^{-\frac{1}{2}}\big(\iv^{n}(\Xv_{1},\Xv_{2},\Yv)-n\Iv\big)\label{eq:MD_Shat_n}               \\
    \bLambda_{n}(x_{1},x_{2}) & \defeq\bSigma_{n}^{-\frac{1}{2}}\Big(\iv\big(x_{1},x_{2},Y(x_{1},x_{2})\big) \nonumber \\
                              & \qquad\qquad\quad -\iv^{(1)}(x_{1}) -\iv^{(2)}(x_{2})+\Iv\Big)\label{eq:MD_Tx1x2} \\
    \xi_{n}                               & \defeq\sum_{x_{1},x_{2}}Q_{1}(x_{1})Q_{2}(x_{2})\EE\Big[\|\bLambda_{n}(x_{1},x_{2})\|^{3}\Big],\label{eq:MD_Beta}                                                               
\end{align}
where $Y(x_{1},x_{2})\sim W(\cdot|x_{1},x_{2})$. From \eqref{eq:MD_CovFinal},
the matrix $\bSigma_{n} - \Vv$ has $O\big(\frac{1}{n}\big)$ entries.   
It follows that whenever $\Vv\succ\bzero$, we
have for sufficiently large $n$ that $\bSigma_{n}\succ\bzero$,
and hence $\bSigma_{n}^{-\frac{1}{2}}$ is well-defined.
\begin{thm}
    \emph{\label{thm:MD_BerryEsseen} (Corollary of \cite[Thm.~2]{CombLatin})}
    Let the input distributions $Q_{1}$ and $Q_{2}$ be given, and consider
    the quantities $(\Xv_{1},\Xv_{2},\Yv)$,
    $\Iv$, $\Vv$, $\widehat{\Sv}_{n}$
    and $\xi_{n}$ respectively defined in \eqref{eq:MD_Distribution},
    \eqref{eq:MD_VecI}, \eqref{eq:MD_CovMatrix}, \eqref{eq:MD_Shat_n}
    and \eqref{eq:MD_Beta} (with $\Uc=\emptyset$). If $\Vv\succ\bzero$,
    then we have for sufficiently large $n$ that
    \begin{equation}
        \Big|\PP\big[\widehat{\Sv}_{n}\in\Ac\big]-\PP\big[\Zv\in\Ac\big]\Big|\le\frac{1}{\sqrt{n}}\frac{K}{\xi_{n}}\label{eq:MD_BerryEsseen}
    \end{equation}
    for any convex, Borel-measurable set $\Ac\subseteq\RR^{3}$,
    where $\Zv\sim N(\bzero,\mathbb{I})$, and $K$
    is a universal constant.
\end{thm}
Obtaining this result from \cite[Thm.~2]{CombLatin} is non-trivial,
and the details are provided in Appendix \ref{sub:MD_BerryEsseen}.

In the discrete setting under consideration, one can show that $\xi_{n}=\Theta(1)$
using the fact that the relevant third moments are finite (e.g.~they
can be uniformly bounded in terms of the alphabet sizes \cite[Appendix D]{MACFinite1}).
Thus, we obtain the desired $O\big(\frac{1}{\sqrt{n}}\big)$ convergence
in \eqref{eq:MD_BerryEsseen}. 

When $\Vv \succ \bzero$, we can use Theorem \ref{thm:MD_BerryEsseen} to bound the probability in
\eqref{eq:MD_pe_bar2} by writing
\begin{align}
    &\PP\big[\iv^{n}(\Xv_{1},\Xv_{2},\Yv)\succ\bgamma\big] \nonumber \\
        &~~= \PP\bigg[ \frac{1}{\sqrt n}\big(\iv^{n}(\Xv_{1},\Xv_{2},\Yv) - n\Iv\big) \succ \frac{1}{\sqrt n}\big(\bgamma - n\Iv\big) \bigg] \label{eq:MD_BerryApp1}\\
        &~~= \PP\bigg[ \frac{1}{\sqrt n}\bSigma_n^{-\frac{1}{2}}\big(\iv^{n}(\Xv_{1},\Xv_{2},\Yv) - n\Iv\big) \in \Ac_n \bigg] \label{eq:MD_BerryApp2} \\
        &~~= \PP\big[ \Zv \in \Ac_n \big] + O\Big(\frac{1}{\sqrt n}\Big) \label{eq:MD_BerryApp3} \\
        &~~= \PP\bigg[ \bSigma_n^{\frac{1}{2}}\Zv \succ \frac{1}{\sqrt n}\big(\bgamma - n\Iv\big) \bigg] + O\Big(\frac{1}{\sqrt n}\Big) \label{eq:MD_BerryApp4}
\end{align}
where \eqref{eq:MD_BerryApp2} follows by defining $\Ac_n$ to be
the image of the rectangular region in \eqref{eq:MD_BerryApp1}
under $\bSigma_n^{-\frac{1}{2}}$, \eqref{eq:MD_BerryApp3} follows
with $\Zv \sim N(\bzero,\mathbb{I})$ from Theorem \ref{thm:MD_BerryEsseen},
and  \eqref{eq:MD_BerryApp4} follows by reversing the step in
\eqref{eq:MD_BerryApp2}.  These steps are similar to \cite[Appendix B]{MACFinite1},
where a Cholesky decomposition is used.

\subsubsection{Singular Dispersion Matrices}

In general, the dispersion matrix $\Vv$ may not have full
rank, in which case Theorem \ref{thm:MD_BerryEsseen} does not directly
apply. We can deal with this case by reducing the problem to a lower
dimension, similarly to \cite[Sec.~VIII-A]{MACFinite1}. The argument
here is slightly more involved, since $n\Vv$ is not necessarily
the exact covariance matrix of $\iv^{n}(\Xv_{1},\Xv_{2},\Yv)$,
due to the additional $O(1)$ term in \eqref{eq:MD_CovFinal}. 

Suppose that $\Vv$ has rank $r<3$, and consider the matrix
$\bSigma_{n}$ in \eqref{eq:MD_SigmaN}. Using an eigenvalue decomposition
along with \eqref{eq:MD_CovFinal}, we obtain
\begin{equation}
    \bSigma_{n}=\bPhi_{n}\bOmega_{n}\bPhi_{n}^{T},\label{eq:MD_EigenDecomp}
\end{equation}
where $\bPhi_{n}$ is a unitary matrix, and $\bOmega_{n}$
is a diagonal matrix whose first $r$ diagonals are $\Theta(1)$,
and whose last $3-r$ diagonals are $O\big(\frac{1}{n}\big)$. Noting
that $\bSigma_{n}$ is the covariance matrix of $\Av_{n}\defeq\frac{1}{\sqrt{n}}\big(\iv^{n}(\Xv_{1},\Xv_{2},\Yv)-n\Iv\big)$,
we see that $\bOmega_{n}$ is the covariance matrix of
$\widetilde{\Av}_{n}\defeq\bPhi_{n}^{T}\Av_{n}$.

Suppose for the time being that $r \ge 1$. From the structure of 
$\bOmega_{n}$, we conclude that
\begin{equation}
    \widetilde{\Av}_{n}=\left[\begin{array}{c}
        \widetilde{\Av}_{n}^{(1)}\\
        \widetilde{\Av}_{n}^{(2)}
    \end{array}\right],
\end{equation}
where $\widetilde{\Av}_{n}^{(1)}$ and $\widetilde{\Av}_{n}^{(2)}$
have dimension $r$ and $3-r$ respectively, and the covariance matrix
of $\widetilde{\Av}_{n}^{(2)}$ has $O\big(\frac{1}{n}\big)$
entries. Since $\bPhi_{n}$ is unitary (i.e.~$\bPhi_{n}\bPhi_{n}^{T}=\mathbb{I}$),
we have $\Av_{n}=\bPhi_{n}\widetilde{\Av}_{n}$,
and hence
\begin{align}
    \Av_{n} & =\bPhi_{n}\left[\begin{array}{c}                                                                                   
    \widetilde{\Av}_{n}^{(1)}\\
    \bzero
    \end{array}\right]+\bPhi_{n}\left[\begin{array}{c}
    \bzero\\
    \widetilde{\Av}_{n}^{(2)}
    \end{array}\right]\label{eq:MD_FinalDecomp0}\\
    & \defeq\bPhi_{n}^{\prime}\widetilde{\Av}_{n}^{(1)}+\widetilde{\bDelta}_{n},\label{eq:MD_FinalDecomp} 
\end{align}
where $\bPhi_{n}^{\prime}$ is obtained from $\bPhi_{n}$
by keeping only the first $r$ columns, and $\widetilde{\bDelta}_{n}$
denotes the second term in \eqref{eq:MD_FinalDecomp0}. Since $\Av_{n}$
has mean zero by construction, we conclude that the same is true of
$\widetilde{\Av}_{n}$, and hence of $\widetilde{\bDelta}_{n}$.
Furthermore, since $\cov[\widetilde{\Av}_{n}^{(2)}]$
has $O\big(\frac{1}{n}\big)$ entries and $\widetilde{\bDelta}_{n}$
is obtained from $\widetilde{\Av}_{n}^{(2)}$ via the unitary (and hence
uniformly bounded) matrix $\bPhi_n$, $\cov[\widetilde{\bDelta}_{n}]$
also has $O\big(\frac{1}{n}\big)$ entries. Thus, Chebyshev's 
inequality implies for any $\delta_{n}>0$ that
\begin{equation}
    \PP\big[\|\widetilde{\bDelta}_{n}\|_{\infty}\ge\delta_{n}\big] \le \frac{\max_{i}\var[\widetilde{\Delta}_{i,n}]}{\delta_n^2} = O\Big(\frac{1}{n\delta_{n}^{2}}\Big),\label{eq:MD_LowerRank4}
\end{equation}
where $\widetilde{\Delta}_{i,n}$ is the $i$-th entry of $\widetilde{\bDelta}$.

We are now in a position to bound the probability appearing in \eqref{eq:MD_pe_bar2}.
The following holds for any $\delta_{n}>0$:
\begin{align}
    & \PP\big[\iv^{n}(\Xv_{1},\Xv_{2},\Yv)\succ\bgamma\big] \nonumber \\
         & =\PP\bigg[\Av_{n}\succ\frac{1}{\sqrt{n}}\big(\bgamma-n\Iv\big)\bigg] \label{eq:MD_LowerRank5a} \\
         & =\PP\bigg[\bPhi_{n}^{\prime}\widetilde{\Av}_{n}^{(1)}+\widetilde{\bDelta}_{n}\succ\frac{1}{\sqrt{n}}\big(\bgamma-n\Iv\big)\bigg]\label{eq:MD_LowerRank5b}                                                                          \\
         & \ge\PP\bigg[\bPhi_{n}^{\prime}\widetilde{\Av}_{n}^{(1)}\succ\frac{1}{\sqrt{n}}\big(\bgamma-n\Iv\big)+\delta_{n}\bone\bigg]-\PP\big[\|\widetilde{\bDelta}_{n}\|_{\infty}\ge\delta_{n}\big]\label{eq:MD_LowerRank5c} \\
         & =\PP\bigg[\bPhi_{n}^{\prime}\widetilde{\Av}_{n}^{(1)}\succ\frac{1}{\sqrt{n}}\big(\bgamma-n\Iv\big)+\delta_{n}\bone\bigg]+O\Big(\frac{1}{n\delta_{n}^{2}}\Big),\label{eq:MD_LowerRank5d}                                 
\end{align}
where the final three steps respectively make use of \eqref{eq:MD_FinalDecomp},
\cite[Lemma 9]{MACFinite1}, and \eqref{eq:MD_LowerRank4}. Since
the entries of $\widetilde{\Av}_{n}^{(1)}$ are shifted
and weighted sums of the entries of $\iv^{n}(\Xv_{1},\Xv_{2},\Yv)$,
and since the corresponding covariance matrix is positive definite
by construction, we can analyze \eqref{eq:MD_LowerRank5d} in the
same way as the case $\Vv\succ\bzero$ (other than
the terms $\delta_{n}\bone$ and $O\big(\frac{1}{n\delta_{n}^{2}}\big)$,
which are handled in the following paragraph). The fact that the resulting
second-order term can be written as $-\sqrt{n}\,\Qsf_{\mathrm{inv}}(\Vv,\epsilon)$
follows in the same way as the i.i.d.~case \cite[p.~894]{MACFinite1}.

The remainder term in \eqref{eq:MD_LowerRank5d} contributes an additive
$O\big(\frac{1}{\sqrt{n}\delta_{n}^{2}}\big)$ term to the expansion
in \eqref{eq:MD_Achievable-2}, whereas the addition of $\delta_{n}\bone$
in the first probability in \eqref{eq:MD_LowerRank5d} contributes
an additive $O(\delta_{n}\sqrt{n})$ term to the expansion. The overall
contribution $O\big(\frac{1}{\sqrt{n}\delta_{n}^{2}}+\delta_{n}\sqrt{n}\big)$
is minimized by $\delta_{n}=\Theta(n^{-\frac{1}{3}})$, yielding $g(n)=O(n^{\frac{1}{6}})$,
as stated following Theorem \ref{thm:MD_MainResult}.

The case $r=0$ (i.e.~$\Vv=\bzero$) is handled similarly; by following the steps of
\eqref{eq:MD_LowerRank5a}--\eqref{eq:MD_LowerRank5d}, we readily 
obtain the following analog of \eqref{eq:MD_LowerRank5d}:
\begin{multline}
    \PP\big[\iv^{n}(\Xv_{1},\Xv_{2},\Yv)\succ\bgamma\big] \\ \ge \openone\bigg\{\bzero\succ\frac{1}{\sqrt{n}}\big(\bgamma-n\Iv\big)+\delta_{n}\bone\bigg\}+O\Big(\frac{1}{n\delta_{n}^{2}}\Big)
\end{multline}
Using this bound, we can obtain \eqref{eq:MD_Achievable-2} (with 
$\beta=0$ and $\Uc=\emptyset$) in the same way as the case $r>0$ by
noting that $\Qinv(\bzero,\epsilon)$ equals the set of 
all vectors with non-negative components. 

\subsection{Proof of Theorem \ref{thm:MD_MainResult} (General Case) \label{sub:MD_PROOF_PART2}}

Here we provide the changes required in the previous subsection to
prove Theorem \ref{thm:MD_MainResult} in full generality. We recall
the definitions of $P_{UX_{1}X_{2}Y}^{\prime}$, $\iv^{\prime}$,
etc.~at the beginning of the section.

\subsubsection{Coded Time-Sharing with $\beta=0$}

In the case that $\beta=0$ but $\Uc\ne\emptyset$, we modify
the constant-composition random-coding ensemble (cf. \eqref{eq:MD_Pxv})
as follows. We let $Q_{U,n}$, $Q_{1,n}$ and $Q_{2,n}$ be (conditional)
types that respectively approximate $Q_{U}$, $Q_{1}$ and $Q_{2}$.
We fix an arbitrary time-sharing sequence $\uv$ with type
$Q_{U,n}$, and generate the $M_{\nu}\defeq e^{nR_{\nu}}$ codewords
of user $\nu=1,2$ independently according to the uniform distribution
on $T_{\uv}^{n}(Q_{\nu,n})$, i.e.~
\begin{equation}
    P_{\Xv_{\nu}|\Uv}(\xv_{\nu}|\uv)=\frac{1}{|T_{\uv}^{n}(Q_{\nu,n})|}\openone\Big\{\xv_{\nu}\in T_{\uv}^{n}(Q_{\nu,n})\Big\}.\label{eq:MD_Pxvu}
\end{equation}
Similarly to \eqref{eq:MD_Distribution}, we define the random variables
\begin{multline}
    (\Xv_{1},\Xv_{2},\Yv,\overline{\Xv}_{1},\overline{\Xv}_{2})\sim P_{\Xv_{1}|\Uv}(\xv_{1}|\uv)P_{\Xv_{2}|\Uv}(\xv_{2}|\uv) \\ \times W^{n}(\yv|\xv_{1},\xv_{2})P_{\Xv_{1}|\Uv}(\overline{\xv}_{1}|\uv)P_{\Xv_{2}|\Uv}(\overline{\xv}_{2}|\uv).
\end{multline}
The procedure described in Section \ref{sub:MD_MOMENTS} for generating
a codeword uniformly over the type class is modified as follows. Let
$\xv$ be an arbitrary element of the conditional type
class $T_{\uv}(\cdot)$. Instead of randomly permuting
the entire sequence $\xv$, a random permutation of the
subsequence $\xv^{(u)}$ corresponding to the indices where
$\uv$ equals $u$ is applied \emph{independently} 
for each value of $u\in\Uc$. Due to this independence, we
can handle the summation 
\begin{align}
    \iv^{n}(\uv,\Xv_{1},\Xv_{2},\Yv) & \defeq\sum_{i=1}^{n}\iv(u_{i},X_{1,i},X_{2,i},Y_{i})                                              \\
        & =\sum_{u}\sum_{i=1}^{nQ_{U}(u)}\iv(u,X_{1,i}^{(u)},X_{2,i}^{(u)},Y_{i}^{(u)})\label{eq:MD_TimeShare2} 
\end{align}
by considering each value of $u\in\Uc$ separately. For the
values of $u$ corresponding to singular dispersion matrices, we can
perform a reduction to a lower dimension as shown following \eqref{eq:MD_EigenDecomp}.
From Theorem \ref{thm:MD_BerryEsseen}, we conclude that each inner
summation in \eqref{eq:MD_TimeShare2} is asymptotically normal with
$O\big(\frac{1}{\sqrt{nQ_{U}(u)}}\big)=O\big(\frac{1}{\sqrt{n}}\big)$
convergence. It follows that the overall sum is also asymptotically
normal with $O\big(\frac{1}{\sqrt{n}}\big)$ convergence. To see this,
we let $\widehat{\Sv}_{1,n}$ and $\widehat{\Sv}_{2,n}$
be asymptotically normal in the sense of \eqref{eq:MD_BerryEsseen},
and let $\Zv_{1}$ and $\Zv_{2}$ be the associated Gaussian random 
variables.  We then have
\begin{align}
    & \PP\big[\widehat{\Sv}_{1,n}+\widehat{\Sv}_{2,n}\in\Ac\big] \nonumber \\ 
    & \quad=\EE\big[\PP\big[\widehat{\Sv}_{1,n}+\widehat{\Sv}_{2,n}\in\Ac\,\big|\,\widehat{\Sv}_{2,n}\big]\big]\label{eq:MD_SumProof1} \\
    & \quad=\EE\big[\PP\big[\Zv_{1}+\widehat{\Sv}_{2,n}\in\Ac\,\big|\,\widehat{\Sv}_{2,n}\big]\big]+O\Big(\frac{1}{\sqrt{n}}\Big)      \\
    & \quad=\PP\big[\Zv_{1}+\widehat{\Sv}_{2,n}\in\Ac\big]+O\Big(\frac{1}{\sqrt{n}}\Big)\label{eq:MD_SumProof3}                                          \\
    & \quad=\PP\big[\Zv_{1}+\Zv_{2}\in\Ac\big]+O\Big(\frac{1}{\sqrt{n}}\Big),\label{eq:MD_SumProof4}                                                     
\end{align}
where \eqref{eq:MD_SumProof4} follows using similar steps to \eqref{eq:MD_SumProof1}--\eqref{eq:MD_SumProof3}.
Using this observation and repeating the analysis of Section \ref{sub:MD_PROOF_SIMPLIFIED}
and \cite{MACFinite1}, we obtain the more general result of Theorem
\ref{thm:MD_MainResult} with $\beta=0$.

\subsubsection{Coded Time-Sharing with $\beta>0$ }

In the case that $\beta>0$, we apply a variant of coded time-sharing
depending on an extended alphabet $\widetilde{\Uc}\defeq\Uc\times\{1,2\}$
and the triplets $(Q_{U},Q_{1},Q_{2})$ and $(Q_{U}^{\prime},Q_{1}^{\prime},Q_{2}^{\prime})$.
Specifically, we define the triplet $(\Qtilde_{U,n},\Qtilde_{1,n},\Qtilde_{2,n})$
with $\Qtilde_{U,n}\in\Pc(\widetilde{\Uc})$,
$\Qtilde_{1,n}\in\Pc(\Xc_{1}\,|\,\widetilde{\Uc})$
and $\Qtilde_{2,n}\in\Pc(\Xc_{2}\,|\,\widetilde{\Uc})$
as follows: 
\begin{align}
    \Qtilde_{U,n}(u,1)           & =\Big(1-\frac{\beta}{\sqrt{n}}\Big)Q_{U}(u) \label{eq:MD_Qtilde1}\\
    \Qtilde_{U,n}(u,2)           & =\frac{\beta}{\sqrt{n}}Q_{U}^{\prime}(u)\label{eq:MD_Qtilde2} \\
    \Qtilde_{\nu,n}(x_{\nu}|u,1) & =Q_{\nu}(x_{\nu}|u)                                           \\
    \Qtilde_{\nu,n}(x_{1}|u,2)   & =Q_{\nu}^{\prime}(x_{\nu}|u)                                  
\end{align}
for $\nu=1,2$. We consider the constant-composition ensemble in \eqref{eq:MD_Pxvu}
with $(\Qtilde_{U,n},\Qtilde_{1,n},\Qtilde_{2,n})$
playing the role of $(Q_{U},Q_{1},Q_{2})$, and we set $\Ptilde_{UX_{1}X_{2}Y,n}\defeq\Qtilde_{U,n}\times\Qtilde_{1,n}\times\Qtilde_{2,n}\times W$.
We follow the same arguments as the case that $\beta=0$ and $\Uc\ne\emptyset$,
but with care taken to handle the fact that some of the time-sharing
values $\widetilde{u}\in\widetilde{\Uc}$ correspond to subsequences
of $\uv$ having length $\Theta(\sqrt{n})$, rather than
$\Theta(n)$. Since we are interested in the limit of large $n$,
we may assume that $1-\frac{\beta}{\sqrt{n}}>0$.

The role of the information density vector $\iv$ is now
played by the quantity
\begin{equation}
    \tilde{\iv}(\tilde{u},x_{1},x_{2},y)\defeq\begin{cases}
    \iv(u,x_{1},x_{2},y) & \tilde{u}=(u,1)\\
    \iv^{\prime}(u,x_{1},x_{2},y) & \tilde{u}=(u,2).
\end{cases} \label{eq:MD_iv_tilde}
\end{equation}
The corresponding mean vector and dispersion matrix with respect to 
$\Ptilde_{UX_{1}X_{2}Y,n}$ are given by
\begin{align}
    \widetilde{\Iv} & \defeq\Big(1-\frac{\beta}{\sqrt{n}}\Big)\Iv+\frac{\beta}{\sqrt{n}}\Iv^{\prime}\label{eq:MD_Itilde}  \\
    \widetilde{\Vv} & \defeq\Big(1-\frac{\beta}{\sqrt{n}}\Big)\Vv+\frac{\beta}{\sqrt{n}}\Vv^{\prime},\label{eq:MD_Vtilde} 
\end{align}
where $\Iv^{\prime}$ and $\Vv^{\prime}$ are defined in 
\eqref{eq:MD_VectorI'}--\eqref{eq:MD_CovMatrix'}.  With
these definitions, the analysis proceeds in the same way 
as the above analysis for $\Uc\ne\emptyset$.  Analogously
to \eqref{eq:MD_TimeShare2} (and using analogous notation), 
the additive $n$-letter extension $\tilde{\iv}^{n}$ of 
\eqref{eq:MD_iv_tilde} admits the decomposition 
\begin{align}
    \tilde{\iv}^{n}(\uv,\Xv_{1},\Xv_{2},\Yv) =\sum_{\tilde{u}}\sum_{i=1}^{n \tilde{Q}_{U,n}(\tilde{u})}\tilde{\iv}(\tilde{u},X_{1,i}^{(\tilde{u})},X_{2,i}^{(\tilde{u})},Y_{i}^{(\tilde{u})}). \label{eq:MD_itilde_n} 
\end{align} 
The inner summations corresponding to $\tilde{u} = (u,2)$
only contain $\Theta(\sqrt{n})$ terms (rather than $\Theta(n)$ terms).
Since the remainder term in the Berry-Esseen theorem decays as the
inverse of the square root of the number of terms, we get $O(n^{\frac{1}{4}})$
in place of $O(\frac{1}{\sqrt{n}})$ in \eqref{eq:MD_BerryApp4}.  

On the other hand, the remainder term in \eqref{eq:MD_LowerRank5d} is
unchanged despite the presence of $\tilde{u}$ values with corresponding
subsequences of length $\Theta(\sqrt{n})$.  To see this, we first write
\begin{align}
    \PP\big[\tilde{\iv}^{n} \succ \bgamma\big] =\PP\bigg[\sum_{\tilde{u}} \frac{1}{\sqrt n}\big( \tilde{\iv}_{\tilde{u}}^{n_{\tilde{u}}} - n_{\tilde{u}} \widetilde{\Iv}_{\tilde{u}} \big) \succ\frac{1}{\sqrt{n}}\big(\bgamma-n\widetilde{\Iv}\big)\bigg], \label{eq:MD_dim_reduc_ts}
\end{align}
where $\tilde{\iv}^{n}$ denotes the left-hand side of \eqref{eq:MD_itilde_n}
(with implicit arguments), 
$\tilde{\iv}_{\tilde{u}}^{n_{\tilde{u}}}$ denotes the inner summation corresponding
to a given $\tilde{u}$ on the right-hand side of \eqref{eq:MD_itilde_n},
$n_{\tilde{u}} \triangleq n\tilde{Q}_{U,n}(\tilde{u})$ denotes the
number of terms in the summation,
and $\widetilde{\Iv}_{\tilde{u}}$ denotes the mean of each summand therein.
These definitions, along with those in \eqref{eq:MD_Qtilde1}--\eqref{eq:MD_iv_tilde}, 
readily yield $\widetilde{\Iv} = \sum_{\tilde{u}} \tilde{Q}_{U,n}(\tilde{u}) \widetilde{\Iv}_{\tilde{u}}$.
Using \eqref{eq:MD_dim_reduc_ts}, one can follow the steps in
\eqref{eq:MD_LowerRank5a}--\eqref{eq:MD_LowerRank5d} and end up with
the same remainder term as \eqref{eq:MD_LowerRank5d}, regardless
of which values of $\tilde{u}$ have corresponding dispersion
matrices that are singular.

Combining the preceding observations and following the steps
of the previous subsections and \cite{MACFinite1}, we obtain
the following condition for $(n,\epsilon)$-achievability:
\begin{equation}
    n\Rv\in n\widetilde{\Iv} - \sqrt{n}\,\Qsf_{\mathrm{inv}}(\widetilde{\Vv},\epsilon)+O\big(n^{\frac{1}{4}}\big)\bone. \label{eq:MD_CondTilde}
\end{equation}
The proof is concluded by substituting \eqref{eq:MD_Itilde}--\eqref{eq:MD_Vtilde}
into \eqref{eq:MD_CondTilde} and using a Taylor expansion of $\Qsf_{\mathrm{inv}}$
(e.g.~see \cite[Lemma 6]{PaperCMAC}) to replace $\Qsf_{\mathrm{inv}}(\widetilde{\Vv},\epsilon)$ by $\Qsf_{\mathrm{inv}}(\Vv,\epsilon)$.

\section{Conclusion}

We have characterized the second-order asymptotics of the
DM-MAC using constant-composition random coding and a combinatorial
Berry-Esseen theorem. Applying an extended version of coded time-sharing,
we have presented a new method for obtaining the derivative (or tangent
vector) terms in the second-order rate region, which first appeared
in \cite{PaperCMAC}. Analogously to the random-coding error exponents
\cite{MACExponent4}, we have observed improved bounds for 
constant-composition random coding compared to i.i.d.~random coding. 
While we focused primarily on unconstrained
channels, our results are directly applicable to discrete channels
with input constraints, thus providing another advantage over i.i.d.
codes.  We have also presented an extension of our main result to the
Gaussian setting via an increasingly fine quantization of the inputs.

A highly challenging open problem is the development of
outer bounds on $\Lc$.  The converse analysis for
the Gaussian MAC with degraded message sets \cite{PaperCMAC} relied on
a reduction from average error to maximal error, but it is well-known
that such a reduction is not possible for the standard MAC \cite{MACMaximal}.  The
``wringing techniques'' used in Ahlswede's derivation of the strong converse 
circumvent this issue \cite{MACStrong2}, but still fail to exhibit
$O\big(\frac{1}{\sqrt n}\big)$ convergence rates to the boundary points,
as is required to get a non-trivial outer bound on $\Lc$.

\appendices

\section{\label{sub:MD_BerryEsseen}Proof of Theorem \ref{thm:MD_BerryEsseen}}

Here we outline the problem studied by Loh \cite{CombLatin} and
state the result that recovers Theorem \ref{thm:MD_BerryEsseen},
adapting the notation therein to be more consistent with ours. 
We write $A_{1}\deq A_{2}$ if the random variables $A_{1}$
and $A_{2}$ have the same distribution. 
 
The ``dimensionality'' in \cite{CombLatin} corresponds to the number
of users of the MAC, so we let it equal $2$.  Let $\pi_{1}(\cdot)$
and $\pi_{2}(\cdot)$ be independent random permutations of $\{1,\cdots,n\}$,
uniformly distributed over the $n!$ possible permutations. For $\nu=1,2$
and $j_{1},j_{2}=1,\cdots,n$,  define the random variables $U_{\nu}(j_{1},j_{2})$
uniformly distributed on $(0,1)$ independently of each other and
of $\pi_{1}(\cdot)$ and $\pi_{2}(\cdot)$, and set
\begin{align}
    B_{\nu}(j_{1},j_{2}) & \defeq\frac{j_{\nu}-U_{\nu}(j_{1},j_{2})}{n},\,\nu=1,2.\label{eq:MD_Xnu} 
\end{align}
The summation of interest is written as follows:
\begin{equation}
    \Sv_{n}\defeq\sum_{j=1}^{n}\fv\Big(\Bv\big(\pi_{1}(j),\pi_{2}(j)\big)\Big), \label{eq:MD_mu_hat}
\end{equation}
where $\Bv(\cdot,\cdot) = [B_1(\cdot,\cdot) ~\, B_2(\cdot,\cdot)]^T$, and 
$\fv(\cdot)$ is a function with a two-dimensional vector argument
and a three-dimensional vector output. 

Using \eqref{eq:MD_Xnu} and the fact that $U_{\nu}(j_{1},j_{2})\in(0,1)$
almost surely, any realization $b_{\nu}$ of $B_{\nu}$ uniquely determines
both the index $j_{\nu}$ and the variable $U_{\nu}$ in the numerator
of \eqref{eq:MD_Xnu}. Thus, overloading the symbol $\fv$,
we write the following for $\bv = [b_1 ~\, b_2]^T$:
\begin{equation}
    \fv(\bv) = \fv(b_{1},b_{2}) = \fv(j_{1},j_{2},u_{1},u_{2}),
\end{equation}
where the final four arguments are deterministically deduced from $\bv$.

We now provide the key definitions needed to state \cite[Thm.~2]{CombLatin}.
The notation here should be treated as being separate from the
rest of this paper for now, but we will shortly see that the definitions
of all re-used symbols are consistent.  The quantities related
to first moments are as follows: 
\begin{align} \allowdisplaybreaks[1] 
    \bmu          &\defeq \frac{1}{n} \EE\big[ \Sv_n \big] \label{eq:BE_mu} \\
    \bmu(j_1,j_2)   &\defeq \EE\big[ \fv(\Bv(j_1,j_2)) \big] \\
    \bmu^{(1)}(j_1)     &\defeq \frac{1}{n}\sum_{j_2=1}^n \bmu(j_1,j_2) \\
    \bmu^{(2)}(j_2)     &\defeq \frac{1}{n}\sum_{j_1=1}^n \bmu(j_1,j_2) \label{eq:BE_mu2}  
\end{align}
The quantities related to second moments are as follows:
\begin{align} \allowdisplaybreaks[1] 
    \bSigma_n          &\defeq \frac{1}{n} \cov\big[ \Sv_n \big] \\
    \widehat{\Sv}_{n}  & \defeq\frac{1}{\sqrt{n}}\bSigma_{n}^{-\frac{1}{2}}\big(\Sv_n - n\bmu\big) \\
    \fv^{(1)}(b_1)     &\defeq \int_0^1 \fv(b_1,b_2)\,db_2 \label{eq:BE_f1} \\
    \fv^{(2)}(b_2)     &\defeq \int_0^1 \fv(b_1,b_2)\,db_1 \\
    \fvrem(b_1,b_2)    &\defeq \fv(b_1,b_2) - \fv^{(1)}(b_1) - \fv^{(2)}(b_2) + \bmu \label{eq:BE_frem} \\
    \Vvf               &\defeq \int_0^1 \int_0^1 \fvrem(b_1,b_2) \fvrem(b_1,b_2)^T \, db_1db_2 \label{eq:BE_Vf} 
\end{align}
Finally, the quantities related to third moments are as follows:
\begin{align} \allowdisplaybreaks[1] 
    \bLambda(j_1,j_2) &\defeq \bSigma_n^{-\frac{1}{2}} \big( \fv(\Bv(j_1,j_2)) - \bmu^{(1)}(j_1) - \bmu^{(2)}(j_2) + \bmu \big) \\
    \xi_n             &\defeq \frac{1}{n^2}\sum_{j_1=1}^n\sum_{j_2=1}^n \EE\big[ \|\bLambda(j_1,j_2)\|^3 \big]. 
\end{align}
We now have the following.
\begin{thm} \label{thm:MD_Loh} \emph{(Combinatorial Berry-Esseen Theorem \cite[Thm.~2]{CombLatin})}
    If $\Vvf \succ \bzero$, then the following holds for sufficiently
    large $n$:
    \begin{equation}
        \Big|\PP\big[\widehat{\Sv}_{n}\in\Ac\big]-\PP\big[\Zv\in\Ac\big]\Big|\le\frac{1}{\sqrt{n}}\frac{K}{\xi_{n}} \label{eq:MD_Loh}
    \end{equation}
    for any convex, Borel-measurable set $\Ac\subseteq\RR^{3}$,
    where $\Zv\sim N(\bzero,\mathbb{I})$, and $K$
    is a universal constant.
\end{thm}

We now show that Theorem \ref{thm:MD_BerryEsseen} is recovered by 
a suitable choice of $\fv$.  
Let $\xv_{1}=(x_{1,1},\cdots,x_{1,n})$ and $\xv_{2}=(x_{2,1},\cdots,x_{2,n})$
be arbitrary sequences having type $Q_{1}$ and $Q_{2}$ respectively,
and define $Y(j_{1},j_{2})\sim W(\cdot|x_{1,j_{1}},x_{2,j_{2}})$ with independence
between different $(j_{1},j_{2})$ pairs.  We set
\begin{equation}
    \fv(j_{1},j_{2},u_{1},u_{2})=\iv\big(x_{1,j_{1}},x_{2,j_{2}},F_{Y(j_{1},j_{2})}^{-1}(u_{1}\oplus u_{2})\big),\label{eq:MD_ChoiceF}
\end{equation}
where $F_{Y(j_{1},j_{2})}^{-1}(u)=\inf\big\{ y\,:\, F_{Y(j_{1},j_{2})}(y)\ge u\big\}$
is the inverse cumulative distribution function (CDF) of $Y(j_{1},j_{2})$, 
and $\oplus$ denotes real addition modulo one.  

We first evaluate the quantities in \eqref{eq:BE_mu}--\eqref{eq:BE_mu2}.
Clearly $U_{1}\oplus U_{2}$ is uniform on $(0,1)$,
and since $F_{Z}^{-1}(U)\deq Z$ for any random variable
$Z$ with CDF $F_{Z}$, it follows that $F_{Y(j_{1},j_{2})}^{-1}(U_{1}\oplus U_{2})\deq Y(j_{1},j_{2})$,
and hence
\begin{equation}
\fv(\Bv(j_1,j_2)) = \fv\big(j_1,j_2,U_1,U_2\big)\deq\iv\big(x_{1,j_{1}},x_{2,j_{2}},Y(j_{1},j_{2})\big), \label{eq:BE_fdeq}
\end{equation}
Since drawing a codeword
uniformly over a type class is equivalent to randomly permuting any
codeword of the given type, it follows that $\Sv_n$ in \eqref{eq:MD_mu_hat} 
has the same distribution as $\iv^{n}(\Xv_{1},\Xv_{2},\Yv)$ 
in \eqref{eq:MD_pe_bar2}.  Using \eqref{eq:BE_fdeq}, the
fact that $\xv_{\nu} \in T^n(Q_{\nu})$ ($\nu=1,2$), and the
definitions of $\iv^{(1)}$ and $\iv^{(2)}$ in
\eqref{eq:MD_ivec_1}--\eqref{eq:MD_ivec_2}, we readily
obtain $\bmu = \Iv$, $\bmu^{(1)}(j_1) = \iv^{(1)}(x_{1,j_1})$,
and $\bmu^{(2)}(j_2) = \iv^{(2)}(x_{2,j_2})$.

Next, we consider the quantities in \eqref{eq:BE_f1}--\eqref{eq:BE_Vf}.
Recalling that the pair $(j_{\nu},u_{\nu})$ is uniquely determined by
$b_{\nu}$ for $\nu=1,2$, we have
\begin{align}
    \fv^{(1)}(b_1) &= \int_0^1 \fv(b_1,b_2)\,db_2 \\ 
        &= \frac{1}{n} \sum_{j_2=1}^{n} \EE\big[\fv(j_1,j_2,u_1,U_2)\big] \label{eq:BE_f1s1} \\
        &= \frac{1}{n} \sum_{j_2=1}^{n} \EE\big[\iv\big(x_{1,j_{1}},x_{2,j_{2}},F_{Y(j_{1},j_{2})}^{-1}(u_{1}\oplus U_{2})\big)\big] \label{eq:BE_f1s2} \\
        &= \frac{1}{n} \sum_{j_2=1}^{n} \EE\big[\iv\big(x_{1,j_{1}},x_{2,j_{2}},Y(j_1,j_2)\big)\big] \label{eq:BE_f1s3} \\
        &= \iv^{(1)}(x_{1,j_1}), \label{eq:BE_f1s4}
\end{align}   
where \eqref{eq:BE_f1s1} follows by interpreting the uniform averaging 
over $b_2\in[0,1]$ as an averaging over $n$ segments of length $\frac{1}{n}$ 
along with an averaging over $u_2$ within each segment, \eqref{eq:BE_f1s2}
follows from \eqref{eq:MD_ChoiceF},  \eqref{eq:BE_f1s3} follows since $u_1 \oplus U_2$
is uniform on $[0,1]$ for any $u_1$, and \eqref{eq:BE_f1s4} follows from
the definition of $\iv^{(1)}$ in \eqref{eq:MD_ivec_1} and the fact that
$\xv_2 \in T^n(Q_2)$.  An identical argument
reveals that $\fv^{(2)}(b_2) = \iv^{(2)}(x_{2,j_2})$, where $j_2$ is 
uniquely determined by $b_2$.

The only remaining quantity whose evaluation is non-trivial is $\Vvf$ in 
\eqref{eq:BE_Vf}.  By writing \eqref{eq:BE_frem} as
\begin{equation}
    \fvrem(b_1,b_2) = (\fv(b_1,b_2) - \bmu) - (\fv^{(1)}(b_1) - \bmu) - (\fv^{(2)}(b_2) - \bmu),
\end{equation}
we can express \eqref{eq:BE_Vf} as the sum of $3^2 = 9$ integrals.  The
desired identity $\Vvf = \Vv$ is obtained by showing that these evaluate to
\begin{align}
    \Vvf &= \Vv^{(12)} + \Vv^{(1)} + \Vv^{(2)} \nonumber \\
         & \qquad - \Vv^{(1)} - \Vv^{(1)} - \Vv^{(2)} - \Vv^{(2)} + \bzero + \bzero \label{eq:BE_Vsum} \\
         &=  \Vv^{(12)} - \Vv^{(1)} - \Vv^{(2)} \\
         &= \Vv,
\end{align}
where $\Vv^{(12)}$, $\Vv^{(1)}$ and $\Vv^{(2)}$ are the three covariance
matrices appearing on the right-hand side of \eqref{eq:MD_CovMatrix} 
(with $\Uc = \emptyset$).  For brevity, we provide details for only one of the 
$9$ terms in \eqref{eq:BE_Vsum}; the others are handled similarly.  We have
\begin{align}
    & \int_0^1\int_0^1(\fv(b_1,b_2) - \bmu)(\fv^{(1)}(b_1) - \bmu)^T \,db_1db_2 \nonumber \\
    & \qquad = \int_0^1\bigg(\int_0^1(\fv(b_1,b_2) - \bmu)\,db_2\bigg)(\fv^{(1)}(b_1) - \bmu)^T \,db_1 \\
    & \qquad = \int_0^1(\fv^{(1)}(b_1) - \bmu)(\fv^{(1)}(b_1) - \bmu)^T \,db_1 \label{eq:BE_IntStep2} \\
    & \qquad = \cov\big[ \iv^{(1)}(X_1) \big] \label{eq:BE_IntStep3} \\
    & \qquad = \Vv^{(1)},
\end{align}
where \eqref{eq:BE_IntStep2} follows from \eqref{eq:BE_f1}, and 
\eqref{eq:BE_IntStep3} follows from \eqref{eq:BE_f1s4} and by
interpreting the integral as an average.

We refer the interested reader to \cite[Sec.~4.5.6]{Thesis} for a list
of identities between quantities appearing in the present paper and the paper 
of Loh \cite{CombLatin}.

\section{\label{sub:MD_GaussConv}Extension to the Gaussian Setting}

\subsection{Evaluation of $\Iv$ and $\Vv$}

The expressions in \eqref{eq:MD_GaussianI}--\eqref{eq:MD_GaussianV12} are
derived from \eqref{eq:MD_VecI} and \eqref{eq:MD_CovMatrix} by forming
an explicit expression for $\iv(x_1,x_2,y)$ in \eqref{eq:MD_bold_i} 
(with $\Uc = \emptyset$ and $Q_1,Q_2 \sim N(0,1)$), performing averaging
in order to obtain $\iv^{(1)}(x_1)$ and $\iv^{(2)}(x_2)$ in \eqref{eq:MD_ivec_1}--\eqref{eq:MD_ivec_2},
and then computing the corresponding means, variances, and covariances.  
For concreteness, we provide a brief outline of this process for the
bottom-right entry of $\Vv$, namely $V_{12}$.

With $Q_1,Q_2 \sim N(0,1)$, the output distribution is $P_Y \sim N(0,1+P_1+P_2)$,
yielding
\begin{equation}
    i_{12}(X_1,X_2,Y) = I_{12} - \frac{Z^2}{2} + \frac{(\sqrt{P_1}X_1 + \sqrt{P_2}X_2 + Z)^2}{2(1+P_1+P_2)}.
\end{equation}
Averaging over $(X_2,Y)$ and $(X_1,Y)$ respectively, we obtain the following:
\begin{align}
    i_{12}^{(1)}(X_1) &= I_{12} + \frac{P_1(X_1^2 - 1)}{2(1+P_1+P_2)} \\
    i_{12}^{(2)}(X_2) &= I_{12} + \frac{P_2(X_2^2 - 1)}{2(1+P_1+P_2)}.  
\end{align} 
Each of the three preceding quantities has mean $I_{12}$.  Using the 
fact that the second and fourth
moments of an $N(0,1)$ random variable are $1$ and $3$ respectively,
the corresponding variances are easily calculated to be $\frac{P_1+P_2}{1+P_1+P_2}$, 
$\frac{P_1^2}{2(1+P_1+P_2)^2}$ and $\frac{P_2^2}{2(1+P_1+P_2)^2}$.
Substituting these into \eqref{eq:MD_CovMatrix} yields 
$V_{12} = \frac{(P_{1}+P_{2})(2+P_{1}+P_{2})+2P_{1}P_{2}}{2(1+P_{1}+P_{2})^{2}}$,
as desired.

\subsection{Derivation of the Achievable Second-order Rate Region}

Recall that the entries of $\Iv$ and $\Vv$ can be written
in the forms given in \eqref{eq:MD_GaussianI}--\eqref{eq:MD_GaussianV12}
respectively. 
The key result used in obtaining \eqref{eq:MD_Gaussian2OR} is 
the following lemma, which states that there exists a sequence
of discrete input distributions $Q_{m1}$ and $Q_{m2}$ of cardinality
$m$ such that the corresponding vector-matrix pair $(\Iv_{m},\Vvm)$
converges to $(\Iv,\Vv)$ (see \eqref{eq:MD_GaussianI}--\eqref{eq:MD_GaussianV12}), with the convergence
$\Iv_{m}\to\Iv$ being exponentially fast in
$m$. This generalizes a result by Wu and Verd\'{u} for the single-user
setting \cite{GaussianCardinality}, and is proved similarly.
\begin{lem} \label{lem:MD_GaussConv}
    There exist sequences
    of discrete input distributions $Q_{m1}$ and $Q_{m2}$ of cardinality
    $m$ with a corresponding matrix-vector pair $(\Iv_{m},\Vvm)$
    defined according to \eqref{eq:MD_VecI} and \eqref{eq:MD_CovMatrix}
    such that (i) $\|\Iv_{m}-\Iv\|_{\infty}\le e^{-\gamma m}$
    for some $\gamma>0$ and sufficiently large $m$, (ii) $\|\Vvm-\Vv\|_{\infty}\to0$,
    and (iii) the third absolute moment of each entry of $\iv(X_{m1},X_{m2},Y)$
    under $Q_{m1}\times Q_{m2}\times Y$ is uniformly bounded in $m$.
\end{lem}
\begin{proof}
    The proof closely follows that of \cite[Thm.~8]{GaussianCardinality},
    so we only explain the differences. We choose $Q_{m1}$ and $Q_{m2}$
    according the Gauss quadrature rule $Q_{g}$ \cite[Sec.~II]{GaussianCardinality},
    which satisfies the property of having the same moments as those 
    of a standard Gaussian random variable up to order $2m-1$ 
    \cite[Thm.~2]{GaussianCardinality}. Since
    $Q_{g}$ converges weakly to $N(0,1)$ \cite{GaussianCardinality}, we immediately obtain
    parts (ii) and (iii) of Lemma \ref{lem:MD_GaussConv}, so it remains
    to prove part (i).
    
    Define $(X_{m1},X_{m2},Y_{m})\sim Q_{m1}\times Q_{m2}\times W$ and
    $(X_{1},X_{2},Y)\sim Q_{1}\times Q_{2}\times W$, where $Q_{1},Q_{2}\sim N(0,1)$.
    Using the identity \cite[Eq.~(15.142)]{Cover}
    \begin{equation}
        I(X_{m1};Y_{m}|X_{m2})=H\big(\sqrt{P_{1}}X_{m1}+Z\big)-H(Z),
    \end{equation}
    we see that the convergence of the first entry of $\Iv_{m}$
    to that of $\Iv$ is precisely that studied in \cite{GaussianCardinality},
    and similarly for the second entry. It remains to study the third
    entry, i.e.~to show that $I(X_{m1},X_{m2};Y_{m})\to I(X_{1},X_{2};Y)$
    exponentially fast. Analogously to \cite[Eq.~(5)]{GaussianCardinality},
    we have
    \begin{align}
        &I(X_{1},X_{2};Y)-I(X_{m1},X_{m2};Y_{m}) \nonumber \\ 
            & =D\big(\sqrt{P_{1}}X_{1}+\sqrt{P_{2}}X_{2}+Z\|\sqrt{P_{1}}X_{m1}+\sqrt{P_{2}}X_{m2}+Z\big) \\
            & \defeq D_{m}.                                                                          
    \end{align}
    Using nearly identical arguments to \cite[Sec.~V]{GaussianCardinality}
    with an {}``optimal'' output distribution of $N(0,1+P_{1}+P_{2})$,
    we obtain analogously to \cite[Eq.~(54)]{GaussianCardinality} that
    \begin{multline}
        D_{m}\le\sum_{k\ge1}\frac{1}{k!}\Big(\frac{P_{1}+P_{2}}{1+P_{1}+P_{2}}\Big)^{k} \\ \times \bigg|\EE\Big[H_{k}\Big(\frac{\sqrt{P_{1}}X_{m1}+\sqrt{P_{2}}X_{m2}}{\sqrt{P_{1}+P_{2}}}\Big)\Big]\bigg|,\label{eq:Chi2}
    \end{multline}
    where $H_{k}$ is the Hermite polynomial of degree $k$ (see \cite[Eq.~(15)]{GaussianCardinality}).
    As shown in \cite{GaussianCardinality}, we obtain the desired exponential
    convergence rate of the mutual information provided that
    the expectation appearing in \eqref{eq:Chi2} is zero for odd values
    of $k$, and also for $k\le2m-1$. For odd values of $k$, we 
    use the same symmetry argument as that of \cite{GaussianCardinality}; since
    the distributions of $X_{m1}$ and $X_{m2}$ are both symmetric, so
    is that of their weighted sum. To handle the remaining values $k\le2m-1$,
    we write
    \begin{equation}
        H_{k}(a+b)=\sum_{i=0}^{k}\sum_{j=0}^{k}c_{ij}a^{i}b^{j}
    \end{equation}
    for some constants $c_{ij}$, which follows since $H_{k}$ has degree
    $k$. By the independence of $X_{m1}$ and $X_{m2}$, the expectation
    $\EE\Big[H_{k}\Big(\frac{\sqrt{P_{1}}X_{m1}+\sqrt{P_{2}}X_{m2}}{\sqrt{P_{1}+P_{2}}}\Big)\Big]$
    depends only on the first $k$ moments of $X_{m1}$ and $X_{m2}$.
    Since the $i$-th moment of $X_{m\nu}$ coincides with the corresponding
    moment of $X_{\nu}\sim N(0,P_{\nu})$ for $i=1,\dotsc,2m-1$ \cite[Thm.~2]{GaussianCardinality},
    we have for $k\le2m-1$ that
    \begin{align}
        &\EE\Big[H_{k}\Big(\frac{\sqrt{P_{1}}X_{m1}+\sqrt{P_{2}}X_{m2}}{\sqrt{P_{1}+P_{2}}}\Big)\Big] \nonumber \\
            & \qquad =\EE\Big[H_{k}\Big(\frac{\sqrt{P_{1}}X_{1}+\sqrt{P_{2}}X_{2}}{\sqrt{P_{1}+P_{2}}}\Big)\Big] \\
            & \qquad =0,\label{eq:Hstep3}                                                                               
    \end{align}
    where \eqref{eq:Hstep3} follows since for any $k$, we have $H_{k}(X)=0$
    under $X\sim N(0,1)$ \cite{GaussianCardinality}.
\end{proof}
We proceed by proving that, analogously to Theorem \ref{thm:MD_MainResult}, 
there exists $g(n)=o(\sqrt{n})$ such that all rate pairs $(R_{1},R_{2})$ satisfying 
\begin{equation}
    \Rv\in n\Iv-\sqrt{n}\,\Qinv(\Vv,\epsilon)+g(n)\bone \label{eq:MD_GaussianGlobal}
\end{equation}
are $(n,\epsilon)$-achievable. This is done by following the analysis 
of Section \ref{sub:MD_PROOF_SIMPLIFIED}: We consider random coding with 
the constant-composition codeword distribution in \eqref{eq:MD_Pxv}, using 
$(Q_{m1},Q_{m2})$ as the input distribution pair. As was done in 
\cite[Thm.~3]{Hayashi}, \cite{TanWiretap}, we set $m=n^{\frac{1}{4}}$. 
By part (i) of Lemma \ref{lem:MD_GaussConv}, we have 
$\|n\Iv_{m}-n\Iv\|_{\infty}\le ne^{-\gamma n^{\frac{1}{4}}}$, 
which behaves as $o(\sqrt{n})$.  Similarly, parts (ii) and (iii)
of Lemma \ref{lem:MD_GaussConv} show that $\Vvm\to\Vv$ and 
the relevant third moments associated with $\iv$ are bounded.  The analysis of Section 
\ref{sub:MD_PROOF_SIMPLIFIED} reveals that the remainder term $g(n)$ in 
\eqref{eq:MD_GaussianGlobal} depends on the alphabet sizes through 
$(|\Xc_{1}|+|\Xc_{2}|+2)\log n$ (see the choice of $d$ following \eqref{eq:MD_gamma}), 
which is again $o(\sqrt{n})$ due to the fact that $|\Xc_{1}|=|\Xc_{2}|=n^{\frac{1}{4}}$.

Finally, using \eqref{eq:MD_GaussianGlobal}, we obtain \eqref{eq:MD_Gaussian2OR} 
using identical steps to Section \ref{sub:MD_PROOF_SIMPLIFIED} with $\beta=0$ 
(see also \cite{NomuraHan}).

\section*{Acknowledgments}

We thank Vincent Tan and Pierre Moulin for helpful discussions. In
particular, the use of quantization arguments for the Gaussian case
in Section \ref{sub:MD_GAUSSIAN} was recommended by Vincent.

\bibliographystyle{IEEEtran}
\bibliography{12-Paper,18-MultiUser,18-SingleUser,35-Other}

\begin{thebibliography}{10}
\providecommand{\url}[1]{#1}
\csname url@samestyle\endcsname
\providecommand{\newblock}{\relax}
\providecommand{\bibinfo}[2]{#2}
\providecommand{\BIBentrySTDinterwordspacing}{\spaceskip=0pt\relax}
\providecommand{\BIBentryALTinterwordstretchfactor}{4}
\providecommand{\BIBentryALTinterwordspacing}{\spaceskip=\fontdimen2\font plus
\BIBentryALTinterwordstretchfactor\fontdimen3\font minus
  \fontdimen4\font\relax}
\providecommand{\BIBforeignlanguage}[2]{{%
\expandafter\ifx\csname l@#1\endcsname\relax
\typeout{** WARNING: IEEEtran.bst: No hyphenation pattern has been}%
\typeout{** loaded for the language `#1'. Using the pattern for}%
\typeout{** the default language instead.}%
\else
\language=\csname l@#1\endcsname
\fi
#2}}
\providecommand{\BIBdecl}{\relax}
\BIBdecl

\bibitem{FanoBook}
R.~Fano, \emph{Transmission of information: A statistical theory of
  communications}.\hskip 1em plus 0.5em minus 0.4em\relax MIT Press, 1961.

\bibitem{ModerateDev}
Y.~Altu\u{g} and A.~B. Wagner, ``Moderate deviations in channel coding,'' 2012,
  http://arxiv.org/abs/1208.1924.

\bibitem{Strassen}
V.~Strassen, ``Asymptotische {A}bsch\"atzungen in {S}hannon's
  {I}nformationstheorie,'' in \emph{Trans. 3rd Prague Conf. on Inf. Theory},
  1962, pp. 689--723, [{E}nglish {T}ranslation:
  http://www.math.wustl.edu/{\string~}luthy/strassen.pdf].

\bibitem{Finite}
Y.~Polyanskiy, H.~V. Poor, and S.~Verd\'{u}, ``Channel coding rate in the
  finite blocklength regime,'' \emph{IEEE Trans. Inf. Theory}, vol.~56, no.~5,
  pp. 2307--2359, May 2010.

\bibitem{Hayashi}
M.~Hayashi, ``Information spectrum approach to second-order coding rate in
  channel coding,'' \emph{IEEE Trans. Inf. Theory}, vol.~55, no.~11, pp.
  4947--4966, Nov. 2009.

\bibitem{MACFinite1}
V.~Y.~F. Tan and O.~Kosut, ``On the dispersions of three network information
  theory problems,'' \emph{IEEE Trans. Inf. Theory}, vol.~60, no.~2, pp.
  881--903, Feb. 2014.

\bibitem{MACFinite2}
E.~MolavianJazi and J.~N. Laneman, ``Simpler achievable rate regions for
  multiaccess with finite blocklength,'' in \emph{IEEE Int. Symp. Inf. Theory},
  Boston, MA, July 2012.

\bibitem{MACFinite3}
Y.~Huang and P.~Moulin, ``Finite blocklength coding for multiple access
  channels,'' in \emph{IEEE Int. Symp. Inf. Theory}, Boston, MA, July 2012.

\bibitem{CombHoeffding}
W.~Hoeffding, ``A combinatorial central limit theorem,'' \emph{Annals Math.
  Stats.}, vol.~22, no.~4, pp. 558--566, 1951.

\bibitem{NomuraHan}
R.~Nomura and T.~S. Han, ``Second-order {S}lepian-{W}olf coding theorems for
  non-mixed and mixed sources,'' in \emph{IEEE Int. Symp. Inf. Theory},
  Istanbul, 2013.

\bibitem{CsiszarBook}
I.~Csisz\'{a}r and J.~K\"{o}rner, \emph{Information Theory: Coding Theorems for
  Discrete Memoryless Systems}, 2nd~ed.\hskip 1em plus 0.5em minus 0.4em\relax
  Cambridge University Press, 2011.

\bibitem{MACExponent4}
Y.~Liu and B.~Hughes, ``A new universal random coding bound for the
  multiple-access channel,'' \emph{IEEE Trans. Inf. Theory}, vol.~42, no.~2,
  pp. 376--386, March 1996.

\bibitem{MACCapacity1}
R.~Ahlswede, ``Multi-way communication channels,'' in \emph{Int. Symp. Inf.
  Theory}, Tsaghkadzor, 1971.

\bibitem{MACCapacity2}
H.~Liao, ``Multiple-access channels,'' Ph.D. dissertation, Dept. Elec. Eng.
  Univ. Hawaii, Honolulu, 1972.

\bibitem{NetworkBook}
A.~{El Gamal} and Y.~H. Kim, \emph{Network Information Theory}.\hskip 1em plus
  0.5em minus 0.4em\relax Cambridge University Press, 2011.

\bibitem{PaperCMAC}
J.~Scarlett and V.~Y.~F. Tan, ``Second-order asymptotics for the {G}aussian
  {MAC} with degraded message sets,'' 2013, submitted to {\em IEEE Trans. Inf.
  Theory } [Online: http://arxiv.org/abs/1310.1197].

\bibitem{DispIC}
S.-Q. Le, V.~Y.~F. Tan, and M.~Motani, ``A case where interference does not
  affect the channel dispersion,'' 2014, http://arxiv.org/abs/1404.0255.

\bibitem{MACFinite5}
E.~Haim, Y.~Kochman, and U.~Erez, ``A note on the dispersion of network
  problems,'' in \emph{IEEE Conv. Elec. Eng. in Israel}, 2012.

\bibitem{CombLatin}
W.~Loh, ``On {L}atin hypercube sampling,'' \emph{Annals of Stats.}, vol.~24,
  no.~5, pp. 2058--2080, 1996.

\bibitem{PaperMAC2_Conf}
J.~Scarlett, A.~Martinez, and A.~{Guill{\'e}n i F\`{a}bregas}, ``Second-order
  rate region of constant-composition codes for the multiple-access channel,''
  in \emph{Allerton Conf. on Comm., Control and Comp.}, Monticello, IL, 2013.

\bibitem{MACExponent2}
R.~Gallager, ``A perspective on multiaccess channels,'' \emph{IEEE Trans. Inf.
  Theory}, vol.~31, no.~2, pp. 124--142, March 1985.

\bibitem{GallagerCC}
------, ``Fixed composition arguments and lower bounds to error probability,''
  \url{http://web.mit.edu/gallager/www/notes/notes5.pdf}.

\bibitem{TanWiretap}
V.~Y.~F. Tan, ``Achievable second-order coding rates for the wiretap channel,''
  in \emph{IEEE Int. Conf. Comm. Sys.}, Singapore, 2012, pp. 65--69.

\bibitem{Cover}
T.~M. Cover and J.~A. Thomas, \emph{Elements of Information Theory}.\hskip 1em
  plus 0.5em minus 0.4em\relax John Wiley \& Sons, Inc., 2001.

\bibitem{MolavianJazi}
E.~MolavianJazi and J.~N. Laneman, ``A finite-blocklength perspective on
  {G}aussian multi-access channels,'' 2014, http://arxiv.org/abs/1309.2343.

\bibitem{CombBolt}
E.~Bolthausen, ``An estimate of the remainder in a combinatorial central limit
  theorem,'' \emph{Prob. Theory and Rel. Fields}, vol.~66, pp. 379--386, 1984.

\bibitem{CombHoChen}
S.~Ho and L.~H.~Y. Chen, ``An ${L}_p$ bound for the remainder in a
  combinatorial central limit theorem,'' \emph{Annals of Probability}, vol.~6,
  no.~2, pp. 231--249, 1978.

\bibitem{CombMultVar}
E.~Bolthausen and F.~G\"{o}tze, ``The rate of convergence for multivariate
  sampling statistics,'' \emph{Annals of Stats.}, vol.~21, no.~4, pp.
  1692--1710, 1993.

\bibitem{CombVonBahr}
B.~{von Bahr}, ``Remainder term estimate in a combinatorial limit theorem,''
  \emph{Prob. Theory and Rel. Fields}, vol.~35, no.~2, pp. 131--139, 1976.

\bibitem{MACMaximal}
G.~Dueck, ``Maximal error capacity regions are smaller than average error
  capacity regions for multi-user channels,'' \emph{Prob. Contr. Inf. Theory},
  vol.~7, pp. 11--19, 1978.

\bibitem{MACStrong2}
R.~Ahlswede, ``An elementary proof of the strong converse theorem for the
  multiple-access channel,'' \emph{Journal Comb. Inf. and Sys. Sci.}, vol.~7,
  no.~3, pp. 216--230, 1982.

\bibitem{Thesis}
J.~Scarlett, ``Reliable communication under mismatched decoding,'' Ph.D.
  dissertation, University of Cambridge, 2014, [Online:
  http://itc.upf.edu/biblio/1061].

\bibitem{GaussianCardinality}
Y.~Wu and S.~Verd\'{u}, ``The impact of constellation cardinality on {G}aussian
  channel capacity,'' in \emph{Allerton Conf. on Comm., Control and Comp.},
  Monticello, IL, 2010.

\end{thebibliography}

\begin{IEEEbiographynophoto}{Jonathan Scarlett}
(S'14) was born in Melbourne, Australia, in 1988. In 2010, he received 
the B.Eng. degree in electrical engineering and the B.Sci. degree in 
computer science from the University of Melbourne, Australia. In 2011, 
he was a research assistant at the Department of Electrical \& Electronic 
Engineering, University of Melbourne.  From October 2011 to August 2014,  
he was a Ph.D. student in the Signal Processing and Communications Group
at the University of Cambridge, United Kingdom. He
is now a post-doctoral researcher with the Laboratory for Information
and Inference Systems at the \'Ecole Polytechnique F\'ed\'erale de Lausanne,
Switzerland.  His research interests are in the areas of information theory, 
signal processing, and high-dimensional statistics. 
\end{IEEEbiographynophoto}

\begin{IEEEbiographynophoto}{Alfonso Martinez}
(SM'11) was born in Zaragoza, Spain, in October 1973. He is currently a Ram\'on y Cajal Research Fellow at Universitat Pompeu Fabra, Barcelona, Spain. He obtained his Telecommunications Engineering degree from the University of Zaragoza in 1997. In 1998--2003 he was a Systems Engineer at the research centre of the European Space Agency (ESA-ESTEC) in Noordwijk, The Netherlands. His work on APSK modulation was instrumental in the definition of the physical layer of DVB-S2. From 2003 to 2007 he was a Research and Teaching Assistant at Technische Universiteit Eindhoven, The Netherlands, where he conducted research on digital signal processing for MIMO optical systems and on optical communication theory. Between 2008 and 2010 he was a post-doctoral fellow with the Information-theoretic Learning Group at Centrum Wiskunde \& Informatica (CWI), in Amsterdam, The Netherlands. In 2011 he was a Research Associate with the Signal Processing and Communications Lab at the Department of Engineering, University of Cambridge, Cambridge, U.K.

His research interests lie in the fields of information theory and coding, with emphasis on digital modulation and the analysis of mismatched decoding; in this area he has coauthored a monograph on ``Bit-Interleaved Coded Modulation''. More generally, he is intrigued by the connections between information theory, optical communications, and physics, particularly by  the links between classical and quantum information theory.
\end{IEEEbiographynophoto}

\begin{IEEEbiographynophoto}{Albert Guill\'en i F\`abregas}
(S'01 -- M'05 -- SM'09) was born in Barcelona, Catalunya, Spain, in 1974. In 1999 he received the Telecommunication Engineering Degree and the Electronics Engineering Degree from Universitat Polit\`ecnica de Catalunya and Politecnico di Torino, respectively, and the Ph.D. in Communication Systems from \'Ecole Polytechnique F\'ed\'erale de Lausanne (EPFL) in 2004.

Since 2011 he has been a Research Professor of the Instituci\'o Catalana de Recerca i Estudis Avançats (ICREA) hosted at the Department of Information and Communication Technologies, Universitat Pompeu Fabra. He is also an Adjunct Researcher at the Department of Engineering, University of Cambridge. He has held appointments at the New Jersey Institute of Technology, Telecom Italia, European Space Agency (ESA), Institut Eur\'ecom, University of South Australia, University of Cambridge where he was a Reader and a Fellow of Trinity Hall, as well as visiting appointments at EPFL, \'Ecole Nationale des T\'el\'ecommunications (Paris), Universitat Pompeu Fabra, University of South Australia, Centrum Wiskunde \& Informatica and Texas A\&M University in Qatar. His specific research interests are in the areas of information theory, communication theory, coding theory, digital modulation and signal processing techniques.     

Dr. Guill\'en i F\`abregas received the Starting Grant from the European Research Council, the Young Authors Award of the 2004 European Signal Processing Conference, the 2004 Best Doctoral Thesis Award from the Spanish Institution of Telecommunications Engineers, and a Research Fellowship of the Spanish Government to join ESA. He is a Member of the Young Academy of Europe. He is a co-author of the monograph book ``Bit-Interleaved Coded Modulation". He is also an Associate Editor of the IEEE Transactions on Information Theory, an Editor of the Foundations and Trends in Communications and Information Theory, Now Publishers and was an Editor of the IEEE Transactions on Wireless Communications (2007-2011).
\end{IEEEbiographynophoto}

\end{document}